\documentclass[preprint,12pt]{elsarticle}

\usepackage{url}
\usepackage{bbold}
\usepackage{multirow}
\usepackage{graphicx}
\usepackage{booktabs}
\usepackage{amsfonts}
\usepackage{amsmath}
\usepackage{enumitem}
\usepackage{orcidlink}
\usepackage[algo2e, linesnumbered, ruled, vlined]{algorithm2e}
\usepackage[misc]{ifsym}
\usepackage{xcolor}
\usepackage{algorithm}
\usepackage{algorithmic}
\usepackage{mathdots}

\usepackage{commath}
\usepackage{adjustbox}
\usepackage{tikz}
\usetikzlibrary{quantikz}
\graphicspath{{images/}}
\usepackage{caption}
\usepackage{subcaption}
\usepackage{comment}
\usepackage{multirow}

\newcommand{\ceil}[1]{\left\lceil #1 \right\rceil}

\begin{document}

\begin{frontmatter}



\title{Quantum Clustering with k-Means: a Hybrid Approach}

    
\author{Alessandro Poggiali\,\orcidlink{0000-0002-1591-7925}}
\author{Alessandro Berti\,\orcidlink{0000-0001-9144-9572}}
\author{Anna Bernasconi\,\orcidlink{0000-0003-0263-5221}}
\author{Gianna~M.~Del~Corso\, \orcidlink{0000-0002-5651-9368 }}
\author{Riccardo Guidotti\, \orcidlink{0000-0002-2827-7613}}

\address{Department of Computer Science, University of Pisa, Largo B. Pontecorvo, Pisa, 56127, Italy}

\begin{abstract}
Quantum computing is a promising paradigm based on quantum theory for performing fast computations.
Quantum algorithms are expected to surpass their classical counterparts in terms of computational complexity for certain tasks, including machine learning.
In this paper, we design, implement, and evaluate three hybrid quantum $k$-Means algorithms, exploiting different degree of parallelism. Indeed, each algorithm incrementally leverages quantum parallelism to reduce the complexity of the cluster assignment step up to a constant cost.
In particular, we exploit quantum phenomena to speed up the computation of distances. The core idea is that the computation of distances between records and centroids  can be executed simultaneously, thus saving time, especially for big datasets.
We show that our hybrid quantum k-Means algorithms can be more efficient than the classical version, still obtaining comparable clustering results.
\end{abstract}




\end{frontmatter}

\section{Introduction}
\label{sec:intro}
Quantum Machine Learning (QML) is the branch of Quantum Computing (QC) that attempts to redesign classical data mining and machine learning algorithms, or their expensive subroutines, to run on a potential quantum computer~\cite{biamonte2017quantum}. 
Many QML algorithms have been recently studied~\cite{schuld2015introduction}.
In this work, we focus on building  quantum versions of $k$-Means~\cite{macqueen1967some}, one of the most famous algorithms used for clustering. 
The $k$-Means clustering algorithm is an unsupervised learning algorithm, and its goal is to find natural groups of elements in a dataset. 
In particular, the elements inside a group are closer to the geometric center, called \textit{centroid} of the group rather than to center of other groups,  according to a specific distance measure. 

Building a quantum version of the $k$-Means algorithm consists of creating a quantum circuit that takes classical data as input and exploits quantum gates to perform the computation, satisfying all the quantum mechanical constraints. The main idea of the algorithms we propose is to incrementally leverage quantum parallelism to \textit{simultaneously} compute  distances between centroids and points in the dataset. In particular, we design three algorithms: the first one, named $q_{1:1}$-$k$-Means, quantizes the single distance computation between a centroid and  a record;  
the second one, called $q_{1:k}$-$k$-Means generalizes the quantum kNN classifier by~\cite{schuld2017implementing} to compute in parallel the distances between a record and all centroids; finally the third algorithm, $q_{M:k}$-$k$-Means, computes simultaneously the distances between all records and all centroids. 

A constraint of quantum computing is that the data need to be normalized according to the 2-norm. However, the straightforward normalization might lead to problems in the distance computation since more than a cluster can be mapped to the same region of the unit sphere. To deal with this problem we adopt as default data preprocessing the Inverse Stereographic Projection (\emph{ISP}), which allows to map $N$-dimensional data into the surface of the sphere in the ($N+1$)-dimensional space. In this way we can better separate the clusters. 

Another important issue arising when implementing quantum algorithms in practice concerns the so-called \textit{post-selection}. This problem happens since, in some quantum algorithms only some branches of the computation, in superposition, lead to meaningful measurements, while other branches lead to measurement that must be discarded. 
Thus, a higher number of executions of the circuits  (the so-called {\em shots})  is required to get accurate results.

In order to assess the $q$-$k$-Means clustering quality, we implement the three versions using the \textsc{qiskit} framework by IBM and test them on both synthetic and real datasets. 
We also make use of the classical $\delta$-k-means algorithm~\cite{kerenidis2019q} to evaluate the quality of our quantum versions.
Our experiments show that the three versions produce comparable clustering results, as long as the number of shots grows proportionally with the number of vectors encoded in the circuits.

This paper is an extended version of the conference paper presented in~\cite{poggiali2022} and is organized as follows.
Section~\ref{sec:related} discusses related works, Section~\ref{sec:setting_the_stage} introduces notations and reviews preliminary notions, Section~\ref{sec:qkmeans} describes the Quantum k-Means.  Section~\ref{sec:experiments} reports the experimental results, and eventually, Section~\ref{sec:conclusion} summarizes our contributions and illustrates future research directions.

\section{Related Works}
\label{sec:related}
The problem of quantum clustering can be addressed in several ways. 
Some studies were inspired by quantum theory. 
For instance, the classical clustering method proposed in~\cite{horn2001algorithm} is based on a physical intuition derived from quantum mechanics. 
In~\cite{otterbach2017unsupervised}, the authors perform clustering by exploiting a well-known reduction from clustering to the Maximum-Cut problem, that is then solved using a quantum algorithm for approximate combinatorial optimization. 
In~\cite{lloyd2013quantum}, the authors present an unsupervised quantum learning algorithm for $k$-Means clustering based on adiabatic quantum computing~\cite{farhi2000quantum}, while in~\cite{xiao2010quantum} the authors propose a quantum-inspired genetic algorithm for $k$-Means clustering. 

The general idea for quantizing classical clustering algorithms is to substitute the most expensive parts in the algorithm with more efficient quantum subroutines. 
For instance, in~\cite{aimeur2006machine} the fidelity distance measure is used for distance computation between each pair of records in the dataset. 
The fidelity is efficiently estimated with a quantum circuit containing only a few gates (two Hadamard gates and a Control-Swap~\cite{nielsen2010quantum}).
In this way, the algorithm can perform clustering directly on quantum states representing the data.
However, the paper lacks a discussion on how the algorithm could deal with classical data, since it assumes that input data are quantum objects. 

As alternative approaches, full quantum routines for clustering have been proposed. 
In~\cite{aimeur2007quantum}, two subroutines based on Grover's search algorithm~\cite{grover1996fast} are used to accelerate classical clustering methods. 
In order to provide a quantum version of $k$-Means, the authors use the subroutines as follows.
First of all,  each record is embedded into a quantum state. Then, for each cluster, the sum of the distances of each state to all others is computed to select the median record, i.e., the record whose total distance to all the others is minimum. This step is implemented with the help of an oracle that calculates the distance between two quantum states. The median record is 
 selected as the new centroid for the cluster.
Unfortunately, this approach cannot be used in practice because an efficient implementation of the oracle is not described. 

More recently,  the authors of ~\cite{wu2022quantum} have proposed another interesting work which exploits Grover's search subroutine to perform clustering. In particular, they provide a quantum $k$-Means algorithm based on Manhattan distance~\cite{thakare2015performance}, i.e. the 2-norm, reaching quadratic speedup with respect to classical $k$-Means.
Furthermore, some algorithms which combine different quantum clustering techniques exist~\cite{benlamine2020quantum}\cite{gong2021quantum}.

Finally,~\cite{kerenidis2019q} propose  a quantum version of $k$-Means (called $q$-Means) that provides an exponential speedup in the number of records of the dataset compared to the classical version. 
Moreover, $q$-Means returns explicit classical descriptions of the final centroids. 
Although $q$-Means looks promising from a practical point of view, the paper discussion is strictly theoretical. The experiments are performed using a classical algorithm ($\delta$-$k$-Means) simulating $q$-Means, instead of a real quantum version. The $q$-Means algorithm has then been exploited in~\cite{kerenidis2021quantum} to create a quantum analog of the spectral clustering~\cite{von2007tutorial}. 

\smallskip
Different from the literature, our work concentrates on practical problems arising when implementing a quantum clustering algorithm, with particular attention to the encoding of classical data in quantum states. 

\section{Setting the Stage}
\label{sec:setting_the_stage}
We keep our paper self-contained by summarizing in this section the key concepts necessary to comprehend our work.
Given a dataset $D = \{\vec{r}_1,\dots,\vec{r}_M\}$ of $M$ records where every record $\vec{r}_i$ is a $N$-dimensional vector, the goal of clustering is to assign each record to one out of $k$ different clusters $\{C_1,\dots,C_k\}$, represented by centroids $\{\vec{c}_1,\dots,\vec{c}_k\}$ respectively, so that similar records, according to a specific distance measure, share the same assignment. In the rest of the paper we will use the following notation: $\hat{n} =  \lceil \log_2N \rceil$, $\hat{k} =  \lceil \log_2k \rceil$ and $\hat{M} =  \lceil \log_2M \rceil$. 

\subsection{k-Means and $\delta$-k-Means}
The $k$-Means algorithm is one of the most famous clustering algorithms~\cite{tan2005introduction}. After randomly choosing $k$ initial centroids\footnote{We adopt and consider the clever random initialization proposed in~\cite{vassilvitskii2006k}.}, the algorithm consists of two  steps  repeated until a certain convergence condition is met. The first step is the \emph{cluster assignment} step: every element in the dataset has to be assigned to its closest centroid according to a specific distance measure. 
As distance, it typically adopts the \emph{Eucledian distance} defined as 
\begin{equation}\label{def:eucl_dist}
d\left( \vec{p},\vec{q}\right)   =\| \vec{p}-\vec{q}\|_2= \sqrt {\sum _{i=1}^{N}  \left( p_{i}-q_{i}\right)^2 },
\end{equation}
where $p$ and $q$ are two $N$-dimensional real vectors.
The second step is the \emph{centroids update} step, where a new cluster center is computed for every cluster to be used as a centroid for the next iteration. 

In this work, we concentrate on the cluster assignment step, whose classical time complexity is $\mathcal{O}(kMN)$ where $k$ is the number of centroids, $M$ is the number of records, and $N$ their dimension.

\begin{algorithm}[t!]
\footnotesize
        \KwIn{$D$ - input data, $k$ - number of clusters}
        \KwOut{$L$ - records to clusters assignment,
        C - centroids}
        \BlankLine
        $C \gets \mathit{initCentroids}(D, k)$ \tcp*{centroid initialization}
        \While{convergence is not achieved}{
            \For(\tcp*[f]{for each record}){$\vec{r} \in D$}{
                $\vec{c} \gets argmin(d(\vec{r}, \vec{c}_j))\, \forall \vec{c}_j \in C$ \tcp*{find nearest centroid}
                $L_\delta(\vec{r}) \gets \{ p : | d^2(\vec{r}, \vec{c}) - d^2(\vec{r}, \vec{c}_p)| \leq \delta \}$ \tcp*{find possible labels}
                $c_j \gets rand(L_\delta(\vec{r}))$ \tcp*{pick a random centroid}
                $C_j \gets C_j \cup \{\vec{r}\}$ \tcp*{assign $\vec{r}$ to cluster $C_j$}
                $L(\vec{r}) \gets j$ 
                \tcp*{assign label $j$ to $\vec{r}$}
            }
            \For(\tcp*[f]{for each centroid}){$j \in [1, k]$}{
                $\vec{c}_j \gets \frac{1}{|C_j|}\sum_{\vec{r} \in C_j} \vec{r}$ \tcp*{update cluster center $\vec{c}_j$}
                }
        }
        \Return $L, C$ \tcp*{return assignments and centroids}
\caption{$\delta$-$k$-Means}
\label{alg:delta-k-Means}
\end{algorithm}

As already mentioned in Section~\ref{sec:related}, a quantum version of $k$-Means, called $q$-Means, is proposed in~\cite{kerenidis2019q}. 
The authors evaluate the effectiveness of $q$-Means comparing it with  $\delta$-$k$-Means,  a  ``quantum-approximation'' of $k$-Means that simulates the quantum calculus that $q$-Means is supposed to execute. 
More precisely, $\delta$-$k$-Means simulates the classical $k$-Means algorithm as performed in a quantum environment.
Since a quantum algorithm can introduce errors due to {\em  decoherence} and {\em noise} present in quantum computers~\cite{preskill2018quantum}, $\delta$-$k$-Means simulates such errors by introducing some noise in both steps of $k$-Means.
However, as later discussed in Section~\ref{sec:qkmeans}, in the quantum versions of $k$-Means we propose in this paper, we keep the second step (i.e., centroids update) classical. For this reason, we consider in the experiments a slightly different version of $\delta$-$k$-Means where we introduce the noise $\delta$ only in the cluster assignment step.  

In Algorithm~\ref{alg:delta-k-Means}, we report the pseudocode of the updated version of the considered $\delta$-$k$-Means.
Let $\vec{c}$ be the centroid closest to the point  $\vec{r}$. 
Then, $\delta$-$k$-Means defines the set of possible labels $L_\delta(\vec{r})$ for $\vec{r}$ as follows:  $$L_\delta(\vec{r}) = \{ p : \abs{d^2(\vec{c}, \vec{r}) - d^2(\vec{c}_p, \vec{r})} \leq \delta \}.$$
When $\delta = 0$, $\delta$-$k$-Means is equivalent to the standard $k$-Means since no uncertainty is included and $L_\delta(\vec{r})$ contains only the centroid computed in line 4. 
On the other hand, a high value of $\delta$ allows $\delta$-$k$-Means to include in $L_\delta(\vec{r})$ also centroids which are distant from the minimum one, bringing more noise in the entire procedure. Indeed,  the assignment rule selects randomly a cluster label from the set $L_\delta(\vec{r})$ (see line 5 in Algorithm~\ref{alg:delta-k-Means}). 
In~\cite{kerenidis2019q} it is proven that if the data are ``well-clusterable'' (see~\cite{kerenidis2019q} for the detailed definition) and the centroids are well separated, the $\delta$-$k$-Means algorithm succeeds assigning the right cluster to most of the points for a suitable value of  $\delta$ depending on the separation of the centroids. 

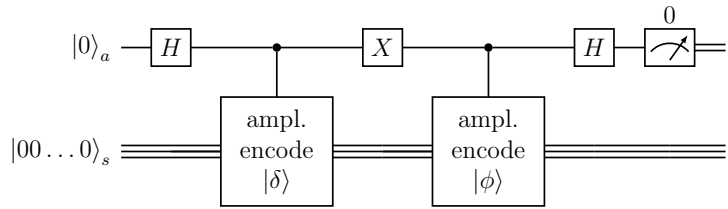
\begin{figure}[t]
    \centering
    \adjustbox{scale=0.8,center}{
    \begin{quantikz}
    \lstick{$\ket{0}_a$} & \gate{H} & \ctrl{1} & \gate{X} & \ctrl{1} & \gate{H} & \meter{0} & \cw\\
    \lstick{$\ket{00\dots0}_s$} & \qwbundle[alternate]{} & \gate{\begin{array}{c}\text{ampl}.\\ \text{encode} \\ \ket{\delta}\end{array}} \qwbundle[alternate]{} & \qwbundle[alternate]{} & \gate{\begin{array}{c}\text{ampl}.\\ \text{encode} \\ \ket{\phi}\end{array}} \qwbundle[alternate]{} & \qwbundle[alternate]{} & \qwbundle[alternate]{} & \qwbundle[alternate]{}\\
    \end{quantikz}
    }
    \caption{Quantum circuit for the Euclidean distance}
    \label{fig:euclideancircuit1}
\end{figure}

\subsection{Quantum Distance Estimate}
Classical information can be encoded in different ways into a quantum state. 
In~\cite{berti2022effects}, several data encoding strategies and quantum distance algorithms are revisited.
The process of encoding input features into the amplitude of a quantum system is called \textit{amplitude encoding}~\cite{schuld2017implementing}.

A distance measure commonly used~\cite{tan2005introduction} in ML and in QML is the Euclidean distance defined in~\eqref{def:eucl_dist}. 
We present here a circuit for computing a quantum Euclidean distance $d(\delta,\phi)$ between two general quantum states $\ket{\delta}$, $\ket{\phi}$ encoded in a register $s$ via amplitude encoding~\cite{schuld2018supervised}.  
To compute a quantum Euclidean distance, we need to use an additional ancilla (i.e., an auxiliary qubit $a$ entangled with the two states $\ket{\delta}$ and $\ket{\phi}$). 
This can be accomplished by first applying an Hadamard gate on the ancilla $a$, and then by loading in the register $s$ the two states $\ket{\psi}$ and $\ket{\delta}$ conditioned on the ancilla.
In this way, the initial state $\ket{0}_a \ket{00 \cdots 0}_s$  evolves in $\frac{1}{\sqrt{2}}(\ket{0}_a \ket{\delta}_s + \ket{1}_a \ket{\phi}_s)$. Eventually, we apply a Hadamard gate on ancilla $a$. 
The corresponding state becomes: $\frac{1}{2} \big(\ket{0}_a (\ket{\delta}_s + \ket{\phi}_s)  + \ket{1}_a (\ket{\delta}_s - \ket{\phi}_s ) \big)$. 
The probability of measuring the ancilla in the state $\ket{0}_a$ is given by $p_a = \frac{1}{4} \| \delta+\phi\|_2^2$ which corresponds to 
\begin{equation} \label{pa}
p_a = 1 - \frac{1}{4}\| \delta-\phi\|_2^2 = 1 -\frac{1}{4} d(\delta,\phi)^2,
\end{equation}
since $\delta$ and $\phi$ are unit vectors. Thus, it is related to the Euclidean distance between the two vectors.
The circuit computing the quantum Euclidean distance between $\delta$ and $\phi$ is illustrated in Figure~\ref{fig:euclideancircuit1}.
  
This procedure can be generalized to compute the quantum Euclidean distance between multiple vectors in superposition. 
Indeed, in~\cite{schuld2017implementing}  Schuld et al. implement a quantum \textit{k-Nearest-Neighbors} (kNN) binary classifier which computes the distances between test and training vectors in parallel. This method can be also extended to solve more complex classification problems, like the facial expression recognition problem in~\cite{mengoni2021facial}.
In the context of supervised learning, kNN is an instance-based classification algorithm that relies on distances. We describe here this quantum kNN procedure because we rely on similar concepts for the definition of our versions of the quantum k-Means for assigning a record to a cluster.

Formally, given a test vector $\widetilde{x}$ to classify in one of two labels $\widetilde{y}\in\{-1,1\}$, the kNN returns the most common class among its $k$ nearest neighbors in the training dataset $D = \{(x^1,y^1),\dots,(x^M,y^M)\}$ where $x^m \in \mathbb{R}^N$ and $y^m \in \{-1,1\}$ for $m=0,\dots\,M-1$. 
The quantum version by Schuld et al.~\cite{schuld2017implementing} exploits the quantum interference property to implement a quantum circuit that can classify a test vector $\widetilde{x}$ in one out of two labels $\widetilde{y}\in\{-1,1\}$. 
The algorithm starts from a quantum state $\ket{D} = \frac{1}{\sqrt{2MC}}\sum_{m=1}^{M}\ket{m}(\ket{0}\ket{\psi_{\widetilde{x}}} + \ket{1}\ket{\psi_{x^m}})\ket{y^m}$ which corresponds to a superposition containing the training data as well as $M$ copies of the new input. 
In particular, $\ket{m}$ is an index register running from $m=0,\dots,M-1$ which denotes the $m$-th training input. 
The second register is a single ancilla qubit whose state is entangled
with the third register encoding the $m$-th training state, $\ket{\psi_{x^m}} = \sum_{i=0}^{N-1}x_i^m \ket{i}$, while the ground state is entangled with the third register encoding the new input $\ket{\psi_{\widetilde{x}}} = \sum_{i=0}^{N-1}\widetilde{x}_i \ket{i}$. The fourth register is a single qubit and corresponds to the labels: it is zero if $y^m=-1$ and one if $y^m=1$. The normalization constant $C$ depends on the preprocessing of the data and we have $C = 1$ if the feature vectors are normalized.

After state preparation, the quantum circuit for classification consists of the following operations: first, a Hadamard gate on the ancilla qubit is applied to 
create interference between the copies of the new input and the training inputs, obtaining the quantum state $$\frac{1}{2\sqrt{M}}\sum_{m=1}^{M}\ket{m}(\ket{0}\ket{\psi_{\widetilde{x}+x^m}} + \ket{1}\ket{\psi_{\widetilde{x}-x^m}})\ket{y^m}$$
where $\ket{\psi_{\widetilde{x}\pm x^m}} = \ket{\psi_{\widetilde{x}}} \pm \ket{\psi_{x^m}}$. Then, a conditional measurement selects the branch where the ancilla is in state $\ket{0}$. The success probability of this postselection is $p_a = \frac{1}{4M}\sum_{m}\|\widetilde{x}+x^m\|^2$. It is more likely to succeed if the collective Euclidean distance of the training set to the new input is small. 
If the data is standardized, this post-selection usually succeeds with a probability of around 0.5~\cite{schuld2017implementing}. Finally, if the conditional measurement is successful, the resulting state is $$\frac{1}{2\sqrt{M p_a}}\sum_{m=1}^{M}\sum_{i=1}^{N}\ket{m}(\widetilde{x_i}+x_i^m)\ket{i}\ket{y^m}\,.$$ Observe that the amplitudes weigh the class qubit $\ket{y^m}$ by the distance of the $m$-th data point to the new input. 
Indeed, in this state  the probability of measuring the class qubit $\ket{y^m}$ in state 0 is $$p(\widetilde{y}=0) = \frac{1}{4M p_a} \sum_{m \vert y^m = 0} \|\widetilde{x} + x^m\|^2\,,$$ which reflects the probability of predicting class -1 for the new input. 
The choice of normalized feature vectors ensures that 
$$
\frac{1}{4M}\sum_{m}\|\widetilde{x}+x^m\|^2 = 1 - \frac{1}{4M}\sum_{m}\|\widetilde{x}-x^m\|^2,
$$ 
and choosing the class with the higher probability therefore implements the required classifier, that actually relates to kNN when setting $k \to M$ and weighing the neighbors by the distance measure. 

\begin{figure}[t]
\centering
    \begin{subfigure}{0.32\textwidth}
    \centering
    \includegraphics[width=0.8\textwidth]{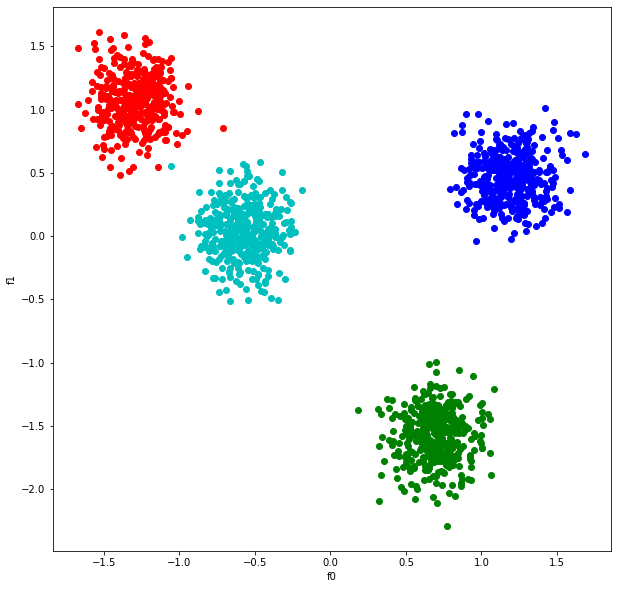}
    \caption{\label{fig:isp-a}}
    \end{subfigure}
    \begin{subfigure}{0.32\textwidth}
    \centering
    \includegraphics[width=0.8\textwidth]{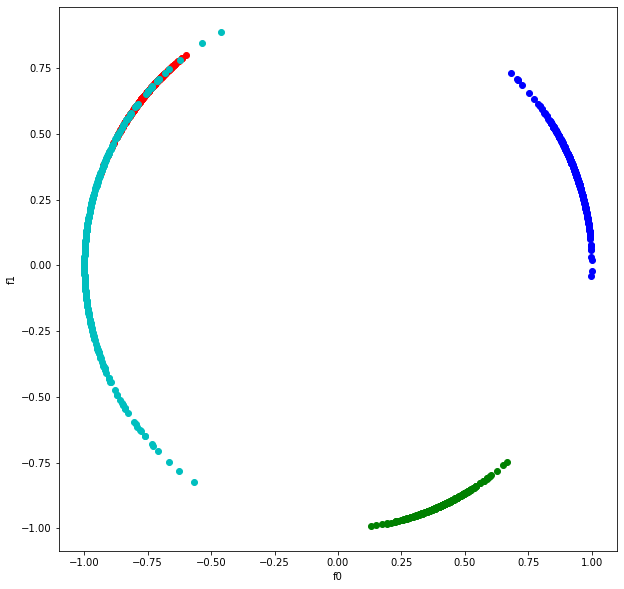}
    \caption{}
    \label{fig:isp-b}
    \end{subfigure}
    \begin{subfigure}{0.32\textwidth}
    \centering
    \includegraphics[width=1\textwidth]{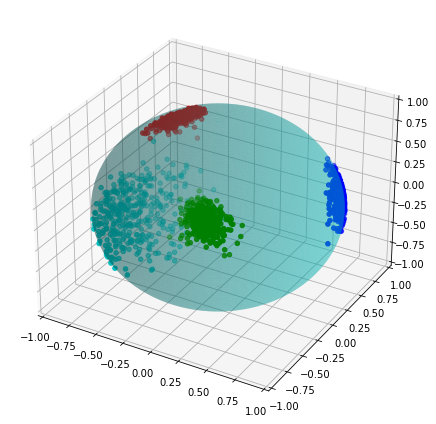}
    \caption{}
    \label{fig:isp-c}
    \end{subfigure}
\caption{Original data (a), normalized (b), normalized with \emph{ISP} (c).}
\label{fig:isp}
\end{figure}

\subsection{Inverse Stereographic Projection}
The methods presented in the previous section allow to estimate the quantum distance only if the data are normalized according with the 2-norm. This implies that a non-normalized dataset should be first transformed in order to have unit-length records. However, the straightforward normalization obtained mapping $(x_1, x_2, \ldots, x_N)\to (x_1/\|x\|, x_2\|x\|, \ldots, x_N\|x\|)$ can lead to problems in the distance computation for the clustering assignment, because more than a cluster can be mapped into the same region of the unit sphere. An example of such situation for $N=2$ is reported in Figure \ref{fig:isp-a}, where the points belong to four well-distinguishable spherical clusters. By normalizing this dataset we would obtain the shape of Figure \ref{fig:isp-b}, where the dataset points are distributed over the unit circle. However, points belonging to different clusters ``overlap'' in the same region of the circle, making it difficult to a clustering algorithm to distinguish them computing the distances between points on the unit circle.
In this work, we propose the Inverse Stereographic Projection (\emph{ISP}) as default data preprocessing. This projection allows to bring $N$-dimensional data into a normalized $(N+1)$-dimensional data, preserving the spherical shape of clusters, but mapping them into different regions. The ISP is a function mapping points on an $N$-dimensional space to $\mathcal{S}^N$, i.e., the sphere in the $(N+1)$-dimensional space. Given a point $x=(x_1, x_2, \ldots, x_N)$, $x\in \mathbb{R}^n$ then $\mbox{ISP}(x)=(\frac{2x_1}{\|x\|+1},\frac{2x_2}{\|x\|+1}, \ldots, \frac{2x_N}{\|x\|+1}, \frac{\|x\|-1}{\|x\|+1})$, $\mbox{ISP}(x)\in \mathcal{S}^N\subset \mathbb{R}^{N+1}$. The dataset of Figure \ref{fig:isp} then becomes the one in Figure \ref{fig:isp-c}, where the shapes of the original clusters are preserved.
Moreover, it is possible to show that we can recover the distances between points in $\mathbb{R}^N$ from the Euclidean distances of the projected point on the sphere. Indeed, if $x, y \in \mathbb{R^N}$ and $X=\mbox{ISP}(x), Y=\mbox{ISP}(y)$ are the images of the points on $\mathcal{S}^N$, we have  
\begin{equation}\label{distance_plane_sphere}
d_N(x, y)^2=\frac{1}{4}(\|x\|^2+1)(\|y\|^2+1) d_{N+1}(X, Y)^2,
\end{equation}
where the subscript tells us the space in which the Euclidean norms are taken. 

This technique has been used also in~\cite{eybpoosh2022applying} to improve clustering accuracy on a set of benchmark datasets. 
 
\begin{algorithm}[t]
\footnotesize
    \setcounter{AlgoLine}{0}
    \KwIn{$D$ - input data,
    $k$ - number of clusters, 
    $t$ - number of quantum shots}
    \KwOut{$L$ - records to clusters assignment,
        $C$ - centroids}
    \BlankLine
    $C \gets \mathit{initCentroids}(D, k)$ \tcp*{centroid initialization}
    \caption{$q$-$k$-means}\label{alg:qk-means}
    \While{centroids do not change}{
        $L, C \gets \mathit{computingCluster}(D, C, k, t)$ \tcp*{quantum computation}
        \For(\tcp*[f]{for each centroid}){$j \in [1, k]$}{
                $\vec{c}_j \gets \frac{1}{|C_j|}\sum_{\vec{r} \in C_j} \vec{r}$ \tcp*{update cluster center $\vec{c}_j$}
                }
    }
    \Return $L, C$ \tcp*{return assignments and centroids}
\end{algorithm}

\section{The q-k-Means Clustering Algorithms}
\label{sec:qkmeans}
In this section we present the $q$-$k$-Means algorithm, which performs clustering on classical data exploiting quantum computation of distances. The general pseudocode of $q$-$k$-Means is reported in Algorithm~\ref{alg:qk-means}. The algorithm takes as input the classical dataset ($D$), the number of cluster ($k$) and the number of repetitions ($t$) for executing the quantum circuits, and returns the final description of clusters and the corresponding centroids.
The main structure is identical to the one of the classical $k$-Means algorithm, where we adopt $k$-Means++~\cite{vassilvitskii2006k} as centroid initialization strategy (line 1). The general idea of this heuristic is to choose $k$ initial centroids as scattered as possible. While the updating of centroids is kept classical (lines 4-5), the step for computing the clusters (line 3), which is the most expensive part of the entire algorithm, is realized through quantum circuits which will be described in the following. 

In particular, three versions of the $q$-$k$-Means algorithm are presented: the first version ($q_{1:1}$-$k$-Means) aims at quantizing the single distance computation between a dataset record and a centroid (Sec.~\ref{sec:5.1}). The second version ($q_{1:k}$-$k$-Means) takes advantage of the quantum kNN classifier by~\cite{schuld2017implementing} to assign the right centroid to a given record according to its closeness (Sec.~\ref{sec:5.2}). Finally, the third version ($q_{M:k}$-$k$-Means) is a further generalization of the previous procedures with the goal of assigning a cluster to every record in a dataset simultaneously (Sec.~\ref{sec:5.3}). 

In order to efficiently load classical data in a suitable quantum state, we employ the FF-QRAM algorithm~\cite{park2019circuit}. Given a classical data structure defined as $D = \left\{ (x_k, p_k) | x_k \in \mathbb{C}, \sum_{k} |x_k|^2 = 1, p_k \in \{0,1\}^{\hat{n}}, 0 \leq k < N\right\}$, where $n=\lceil \log_2(N)\rceil$, this algorithm is able to encode $D$ into a quantum format referred to as quantum database (QDB) and expressed as $\ket{\psi} = \sum_{k=0}^{N-1} x_k \ket{p_k}$. This is done by repeating the execution of a quantum circuit consisting of $\mathcal{O}(n)$ qubits and $\mathcal{O}(Nn)$ \emph{flip}-\emph{flop} operations for a certain number of times. The repetitions are necessary to post-select (i.e., to select only the meaningful branches of computation) the correct outcome for encoding continuous data as probability amplitudes. 
Therefore, $q$-$k$-Means algorithm requires two post-selections: one needed for the kNN quantum classifier, and one needed for the FF-QRAM storing algorithm. 
These post-selections can have a huge impact on the overall algorithm performance, since they imply a high number of circuit executions to get accurate results. 
We discuss this problem in Section~\ref{sec:5.4} along with possible solutions.

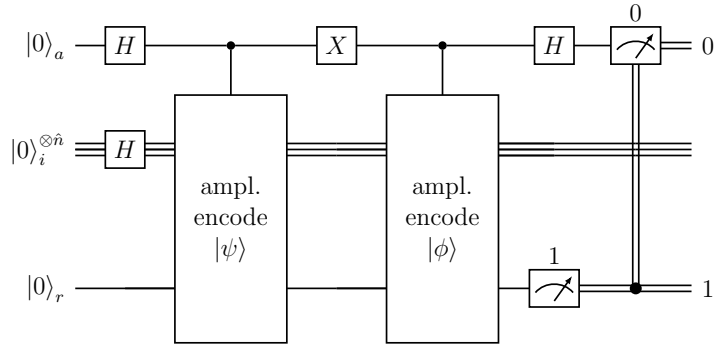
\begin{figure}[t]
\centering
\adjustbox{scale=0.8,center}{
\begin{quantikz}
\lstick{$\ket{0}_a$} & \gate{H} & \ctrl{1} & \gate{X} & \ctrl{1} & \gate{H} & \meter{0} & \cw \rstick{0}\\
\lstick{$\ket{0}_i^{\otimes \hat{n}}$} & \gate{H} \qwbundle[alternate]{} & \gate[wires=2]{\begin{array}{c}\text{ampl}.\\ \text{encode} \\ \ket{\psi}\end{array}} \qwbundle[alternate]{} & \qwbundle[alternate]{} & \gate[wires=2]{\begin{array}{c}\text{ampl}.\\ \text{encode} \\ \ket{\phi}\end{array}} \qwbundle[alternate]{} & \qwbundle[alternate]{}  \qwbundle[alternate]{} & \qwbundle[alternate]{} & \qwbundle[alternate]{}\\
\lstick{$\ket{0}_r$} & \qw & \qw & \qw & \qw & \meter{1} & \cwbend{-2}& \cw \rstick{1}\\
\end{quantikz}
}
\caption{\textsc{QC1}: quantum Euclidean distance with FF-QRAM.}
\label{fig:euclidiancircuit2}
\end{figure}

\begin{algorithm}[t]
\footnotesize
\caption{computingCluster1}\label{alg:q1}
    \setcounter{AlgoLine}{0}
    \KwIn{$D$ - input data,
    $C$ - initial centroids
    $k$ - \# of clusters, 
    $t$ - \# of quantum shots}
    \KwOut{$L$ - records to clusters assignment,
    C - centroids} 
    \BlankLine
    \For(\tcp*[f]{for each record}){$\vec{r} \in D$}{
        $v \gets \infty$\tcp*{init distance}
        \For(\tcp*[f]{for each centroid}){$ j \in [1,k]$}{
            $\#\ket{0}_a \gets$ \textsc{QC1($t$,$\vec{r}$,$\vec{c}_j$)} \tcp*{quantum circuit executed $t$ times}
            $d \leftarrow \sqrt{4 - 4\left(\frac{\#\ket{0}_a}{t'}\right)}$ \tcp*{Euclidean distance estimation}
            \If(\tcp*[f]{if closer to current centroid}){$d \leq v$}{ 
                $v \gets d$\tcp*{update current distance}
                $C_j \gets C_j \cup \{\vec{r}\}$ \tcp*{assign $\vec{r}$ to cluster $C_j$}
                $L(\vec{r}) \gets j$ 
            \tcp*{assign label $j$ to $\vec{r}$}
            }
        }
    }
    \Return $L, C$ \tcp*{return assignments and centroids}
\end{algorithm}

\subsection{One Record vs One Centroid: $q_{1:1}$-$k$-Means}
\label{sec:5.1}
In this version we \textit{make quantum} the computation of the distance between two records. This is similar to what some previous works have proposed~\cite{aimeur2006machine}, but here we give a practical implementation of the entire algorithm.

In particular, we deal with $N$-feature vectors, and we use the FF-QRAM algorithm to amplitude encode each vector. Once data have been loaded, we build a quantum circuit that computes the distance between a single pair of vectors. In the $k$-Means context, we compute distances between every record and every centroid in order to assign a cluster label to the record in the dataset. 
This can be accomplished by using the quantum circuit shown in Figure~\ref{fig:euclideancircuit1} that computes the Euclidean distance between two quantum states generated by the amplitude encoding technique. 

Figure~\ref{fig:euclidiancircuit2} illustrates the whole quantum circuit (\textsc{QC1}).
QC1 can compute the Euclidean distance simultaneously between the features of two vectors.
The $\hat{n}$-qubit register $\ket{i}$ is the index register for the $N$ features of each vector. 
It consists of $\hat{n}=  \lceil \log_2N \rceil$ qubits which control the rotation of a qubit $\ket{r}$.

It is important to underline that to get a proper estimate of the Euclidean distance, we need to repeat the execution of this circuit a certain number of times $t$. 
From equation \eqref{pa}, we can estimate the Euclidean distance as $d(\psi,\phi) = \sqrt{4 - 4\left(\frac{\#\ket{0}_a}{t'}\right)}$, where $\#\ket{0}_a$ is the number of times the ancilla qubit is measured in the state $\ket{0}$ and $\frac{\#\ket{0}_a}{t'}$ is used as an estimate of $p_a$. 
Also the FF-QRAM encoding procedure requires a post-selection on the qubit $\ket{r}$ (see~\cite{park2019circuit, berti2022effects} for more details). 
Thus, $\#\ket{0}_a$ is the number of times the outcome for ancilla is 0 after having post-selected the result where qubit $\ket{r}$ was in 1. 
Notice that, here, $t'<t$ is not the total number of repetitions, but the number of times the post-selection on $\ket{r}$ is successful.

Algorithm~\ref{alg:q1} describes the pseudocode of $\mathit{computingCluster1}$, the proposed procedure for cluster computation, where the quantum Euclidean distance is computed between each record $\vec{r}$ and each centroid $\vec{c}_j$. In particular, the quantum distance computation is repeated for each of the $Mk$ pairs of vectors, where $M$ is the number of records in the dataset and $k$ is the number of centroids (lines 1--5). Then, the algorithm assigns (line 6--9) $\vec{r}$ to the cluster $C_j$  if $d(\vec{r}, \vec{c}_j)= \min _{\kappa\in [1, k]} d(\vec{r}, \vec{c}_{\kappa})$. 

The whole procedure improves the \textit{cluster assignment step} with respect to the classical $k$-Means by a factor $N$ if we do not consider the QRAM preparation. 
Leveraging quantum parallelism, we can compute the Euclidean distance between each $N$-features record and a centroid simultaneously.
In particular, the overall complexity of the \textit{cluster assignment step} is $\mathcal{O}(Mk)$ per circuit execution, where $M$ is the number of records and $k$ is the number of centroids, plus the cost of the QRAM preparation. Note that one can use Equation~\eqref{distance_plane_sphere} to get an estimate of the original Euclidean distances between records and centroids from the distances of their projections measured with the quantum circuit. 

\begin{figure}[t]
\centering
\adjustbox{scale=0.8,center}{
\begin{quantikz}
\lstick{$\ket{0}_a$} & \gate{H} & \ctrl{1} & \gate{X} & \ctrl{1} & \gate{H} & \meter{0} & \cw & \ocwbend{3} & \cw \rstick{0}\\
\lstick{$\ket{0}_i^{\otimes \hat{n}}$} & \gate{H} \qwbundle[alternate]{} & \gate[wires=2]{\begin{array}{c}\text{ampl}.\\ \text{encode} \\ \text{record} \\ \ket{\psi}\end{array}} \qwbundle[alternate]{} & \qwbundle[alternate]{} & \gate[wires=3]{\begin{array}{c}\text{ampl}.\\ \text{encode} \\ \text{sup. cent.} \\ \ket{\phi} \end{array}} \qwbundle[alternate]{} & \qwbundle[alternate]{} & \qwbundle[alternate]{} & \qwbundle[alternate]{} & \qwbundle[alternate]{} & \qwbundle[alternate]{} \\
\lstick{$\ket{0}_r$} & \qw & \qw & \qw & \qw & \meter{1} & \cwbend{-2}& \cw & \cw & \cw \rstick{1}\\
\lstick{$\ket{0}_k^{\otimes \hat{k}}$} & \gate{H} \qwbundle[alternate]{} & \qwbundle[alternate]{} & \qwbundle[alternate]{} & \qwbundle[alternate]{} & \qwbundle[alternate]{} & \qwbundle[alternate]{} & \qwbundle[alternate]{}  & \meter{}\qwbundle[alternate]{} & \cw\\
\end{quantikz}
}
\caption{\textsc{QC2}: quantum kNN classifier with FF-QRAM.}
\label{fig:knnffqram}
\end{figure}
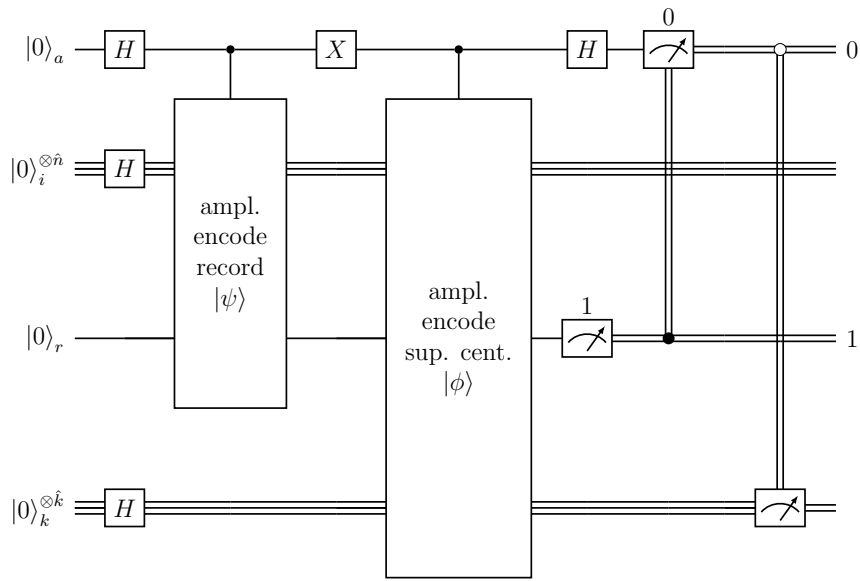

\begin{algorithm}[t]
\footnotesize
\caption{computingCluster2}\label{alg:q2}
    \setcounter{AlgoLine}{0}
    \KwIn{$D$ - input data,
    $C$ - initial centroids
    $k$ - \# of clusters, 
    $t$ - \# of quantum shots}
    \KwOut{$L$ - records to clusters assignment,
    C - centroids} 
    \BlankLine
    \For(\tcp*[f]{for each record}){$\vec{r} \in D$}{
        $\{\#\ket{j}_k\}_{j=0 \cdots \hat{k}-1} \gets$ \textsc{QC2($t$,$\vec{r}$,$C$)} \tcp*{quantum circuit executed $t$ times}
        $\hat{\jmath} \gets argmax(\{\#\ket{j}_k\}_{j=0 \cdots \hat{k}-1})$
        \tcp*{most frequently label observed}
        $C_{\hat{\jmath}} \gets C_{\hat{\jmath}} \cup \{\vec{r}\}$ \tcp*{assign $\vec{r}$ to cluster $C_{\hat{\jmath}}$}
        $L(\vec{r}) \gets \hat{\jmath}$ 
        \tcp*{assign label $\hat{\jmath}$ to $\vec{r}$}
    }
    \Return $L, C$ \tcp*{return assignments and centroids}
\end{algorithm}

\subsection{One Record vs $k$ Centroids: $q_{1:k}$-$k$-Means}
\label{sec:5.2}
Rather than computing the distances between vectors using quantum circuits, and use these distances to classically assign cluster label to every record, this quantum version takes advantage of the quantum kNN. In the clustering context, the training set used in the kNN algorithm corresponds to our set of current centroids while the test vectors to classify correspond to the records belonging to the original dataset. In~\cite{khan2019k}, the authors exploit the quantum kNN to assign one out of two clusters to an input record. Repeating this procedure for all dataset records and for all pair of centroids, they provide a hybrid version of the $k$-Means algorithm. Here, we are able to assign the right cluster label to a record simultaneously for an arbitrary number of clusters. The circuit from~\cite{schuld2017implementing} is able to classify one test vector at a time in one out of two classes. For a practical implementation what we need to do is to generalize this quantum procedure in order to assign to every record one of the $k$ cluster labels available. Thanks to the FF-QRAM algorithm, we can extend the original kNN quantum circuit to directly manage the multiclass case by employing $\hat{k} = \ceil{\log_2k}$ qubits $\ket{\kappa}$ indexing $k$ different cluster labels. These qubits are initially put in an equal superposition by Hadamard gates. In this way, when we encode centroids we have to put them in entanglement also with qubits $\ket{\kappa}$. This means that, when we apply the controlled rotations on qubit $\ket{r}$ to encode centroids values, we need to use also $\ket{\kappa}$ as control qubits to specify the cluster they belong to. On the other hand, the rotations for the record values are not controlled by the label qubits. 

The general quantum circuit implementing this multiclass classification (\textsc{QC2}) is depicted in Figure~\ref{fig:knnffqram}. The first amplitude encodes the unlabeled record in the branch where the ancilla qubit $\ket{a}$ is 0, obtaining the state $\ket{\psi}$, and then we amplitude encode all the centroid vectors ($\ket{\phi}$) along with their cluster labels in the branch where the ancilla is 1. The final Hadamard gate on $\ket{a}$ has the same effect of the one in the original kNN classifier, which interferes the $k$ copies of the unlabeled record with all centroids. After having post-selected the qubits $\ket{r}$ and $\ket{a}$, the cluster label assigned to the considered record is obtained measuring the qubits $\ket{\kappa}$.

The pseudo-code of the $\emph{computingCluster2}$ procedure is given in Algorithm~\ref{alg:q2}. For every record in the dataset, it executes the \textsc{QC2} circuit $t$ times, to get a probability distribution of the cluster labels to assign to the current record (line 2). After having selected the most frequent label in the distribution (line 3), the procedure continues with the actual cluster assignment steps (lines 4-5).

Note that, the assignment of record to clusters needs to load all centroids in the circuit even in the case some of them do not change between iterations. 
The proposed procedure realizes the $q_{1:k}$-$k$-Means algorithm which improves the \textit{cluster assignment step} with respect to the classical $k$-Means by a factor $Nk$. 
The overall complexity of the \textit{cluster assignment step} is $\mathcal{O}(M)$ per circuit execution, where $M$ is the number of records, plus the cost of the QRAM preparation. Note that in this case we cannot employ formula~\eqref{distance_plane_sphere} because the circuit provides directly the cluster assignment without explicitly showing the distances.

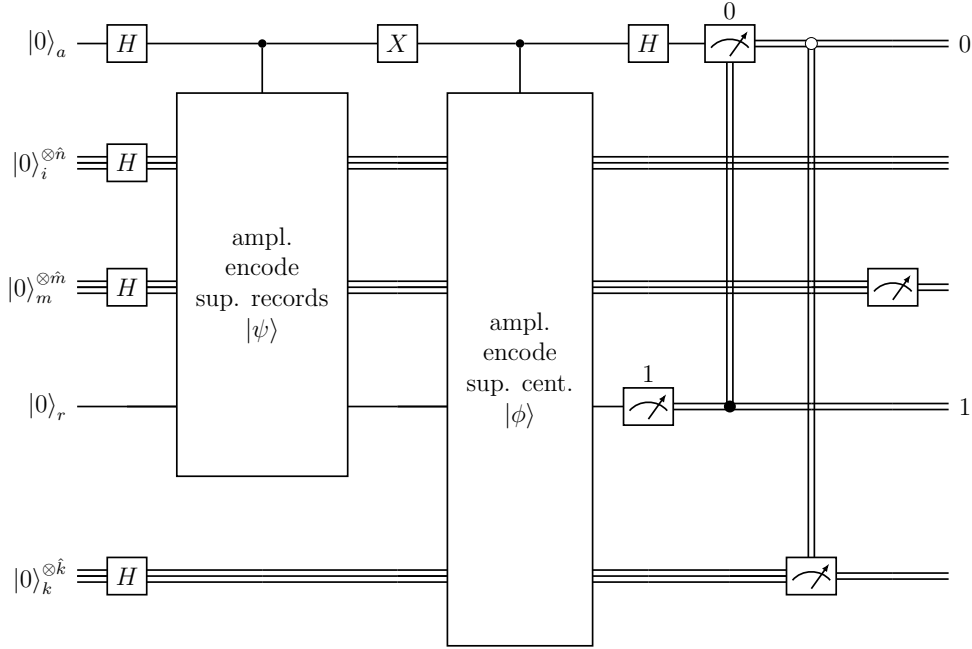
\begin{figure}[t]
\centering
\adjustbox{scale=0.8,center}{
\begin{quantikz}
\lstick{$\ket{0}_a$} & \gate{H} & \ctrl{1} & \gate{X} & \ctrl{1} & \gate{H} & \meter{0} & \ocwbend{4} & \cw & \cw \rstick{0}\\
\lstick{$\ket{0}_i^{\otimes \hat{n}}$} & \gate{H} \qwbundle[alternate]{} & \gate[wires=3]{\begin{array}{c}\text{ampl}.\\ \text{encode} \\ \text{sup. records} \\ \ket{\psi} \end{array}}\qwbundle[alternate]{} & \qwbundle[alternate]{} & \gate[wires=4]{\begin{array}{c}\text{ampl}.\\ \text{encode} \\ \text{sup. cent.} \\ \ket{\phi} \end{array}} \qwbundle[alternate]{} & \qwbundle[alternate]{} & \qwbundle[alternate]{} & \qwbundle[alternate]{} & \qwbundle[alternate]{} & \qwbundle[alternate]{} \\
\lstick{$\ket{0}_{m}^{\otimes \hat{m}}$} & \gate{H} \qwbundle[alternate]{} & \qwbundle[alternate]{} & \qwbundle[alternate]{} & \qwbundle[alternate]{} & \qwbundle[alternate]{} & \qwbundle[alternate]{} & \qwbundle[alternate]{}  & \meter{}\qwbundle[alternate]{} & \cw\\
\lstick{$\ket{0}_r$} & \qw & \qw & \qw & \qw & \meter{1} & \cwbend{-3}& \cw & \cw & \cw \rstick{1}\\
\lstick{$\ket{0}_k^{\otimes \hat{k}}$} & \gate{H} \qwbundle[alternate]{} & \qwbundle[alternate]{} & \qwbundle[alternate]{} & \qwbundle[alternate]{} & \qwbundle[alternate]{} & \qwbundle[alternate]{} &  \meter{}\qwbundle[alternate]{} & \cw & \cw\\
\end{quantikz}
}
\caption{\textsc{QC3}: quantum circuit for cluster assignment.}
\label{fig:finalcircuit}
\end{figure}

\begin{algorithm}[t]
\footnotesize
\caption{computingCluster3}\label{alg:q3}
    \setcounter{AlgoLine}{0}
    \KwIn{$D$ - input data,
    $C$ - initial centroids
    $k$ - \# of clusters, 
    $t$ - \# of quantum shots}
    \KwOut{$L$ - records to clusters assignment,
    C - centroids} 
    \BlankLine
    $\{\#\ket{v}_m\ket{j}_k\}_{v=0 \cdots \hat{m}-1, j=0 \cdots \hat{k}-1} \gets$ \textsc{QC3($t$,$D$,$C$)} \tcp*{quantum circuit executed $t$ times}
    \For(\tcp*[f]{for each record}){$ v \in [0,\hat{m}-1]$}{
        $\hat{\jmath} \gets argmax(\{\#\ket{v}\ket{j}_k\}_{j=0 \cdots \hat{k}-1})$
        \tcp*{most frequently label observed}
        $C_{\hat{\jmath}} \gets C_{\hat{\jmath}} \cup \{\vec{r_v}\}$ \tcp*{assign $\vec{r_v}$ to cluster $C_{\hat{\jmath}}$}
        $L(\vec{r_v}) \gets \hat{\jmath}$ 
        \tcp*{assign label $\hat{\jmath}$ to $\vec{r_v}$}
    }
    \Return $L, C$ \tcp*{return assignments and centroids}
\end{algorithm}

\subsection{$M$ Records vs $k$ Centroids: $q_{M:k}$-$k$-Means}
\label{sec:5.3}
The main limitation of the previous quantum version of the cluster assignment step is that we have to create $M$ different circuits where every circuit differs from the others only by the presence of a different record encoded. This can have a significant impact on the overall performance, especially for very big dataset. Ideally, it would be convenient if we could manage everything in one single quantum circuit, loading centroid vectors only once per iteration. Actually, this can be achieved by adding another addressing level: we extend again the FF-QRAM by \textit{doubling} it a certain number of times so that we can load the records and the centroids in one single circuit. In other words, in one single circuit we are now able to perform the cluster assignment step, assigning to the right cluster every record in the dataset. 

We name this expansion \textit{FF-QRAM doubling}: it consists in adding another set of qubits, $\ket{m}$, which perform another addressing level. 
In particular, every configuration of the basis state of $\ket{m}$ has the task of indexing a specific record assignment to a cluster. 
Thus, the final effect is the same as a doubling of the circuit \textsc{QC2} for a number of time that is equal to the number of records $M$. Hence, the register $\ket{m}$ has size equal to $\hat{m} = \ceil{\log_2M}$. In Figure~\ref{fig:finalcircuit} we show the circuit \textsc{QC3} for the quantum cluster assignment. Unlike \textsc{QC2}, here $\ket{\psi}$ encode all the record vectors instead of only one. Therefore, the rotations on $\ket{r}$ for the record vectors are controlled also by the new $\ket{m}$ qubits, while the centroid vectors are not controlled by them. This ensure that centroids are loaded only once. 

The procedure for this final version is reported in Algorithm~\ref{alg:q3} and the final quantum algorithm using this procedure is called $q_{M:k}$-$k$-Means. We do not have external classical for-loop anymore so we have only one quantum circuit per iteration. 
After having executed the quantum circuit \textsc{QC3} for a certain number of shots (line 1), the actual assignment of every record to a cluster is extrapolated from the probability distribution (lines 2-5). This algorithm improves the \textit{cluster assignment step} with respect to the classical $k$-Means by a factor $MNk$. 
In particular, the overall complexity of the \textit{cluster assignment step} is $\mathcal{O}(1)$ per circuit execution, plus the cost of the QRAM preparation.

\subsection{Reducing the Number of Shots}
\label{sec:5.4}
Our three quantum versions of the cluster assignment step, base their effectiveness on the fact that they execute a quantum circuit a certain number of times, called \textit{repetitions} or \textit{shots}. 
Several repetitions are needed because two post-selections of qubits are required: one on the ancilla qubit $\ket{a}$ that produces a state with amplitudes depending on the distances between the encoded vectors, and one on the register qubit $\ket{r}$ required for the FF-QRAM algorithm. 
Since the cluster assignment we obtain depends on how well we distinguish results on the various measurements, we have to choose a proper number of shots to gather enough statistical information and avoid poor results. On the other hand, a too high number of repetitions could lead towards a slower algorithm. 

\begin{figure}[t]
    \centering
    \includegraphics[width=0.7\textwidth]{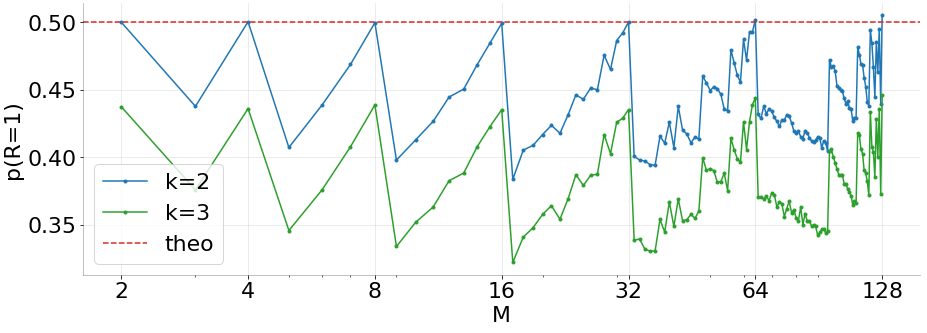}
    \caption{Post-selection probability on $\ket{r}$ varying $M$}
    \label{fig:postprobR}
\end{figure}

In the quantum kNN circuit the post-selection probability on the ancilla qubit is $P(\ket{a}=0) \approx 0.5$, when data are properly standardized~\cite{schuld2017implementing}. This means that half of the measurement results must be discarded. Concerning the post-selection probability for $\ket{r}$, a result in~\cite{de2020circuit} states that, given a data organized as $d=\{x_k,p_k\}_{k=0}^{N-1}$, the FF-QRAM post-selection success probability is $P(\ket{r}=1)=\frac{1}{2^n}$, where $n$ is the number of qubits used for indexing the $N$ different values. This means that FF-QRAM can have an exponential cost in the worst case: if the number $n$ of qubits used for indexing the $N$ values increases, then the post-selection probability $P(\ket{r}=1) = \frac{1}{2^n}$ approaches 0.
In the specific case of the $q$-$k$-Means algorithm, the data $d$ correspond to a single dataset record. However, what $q_{1:k}$-$k$-Means and $q_{M:k}$-$k$-Means algorithms do, is to put into the respective circuits a high number of normalized vectors with the help of the different levels of addressing (for records and centroids). It is worth asking whether, after these generalizations, the theorem still holds. What happens in practice is that if we keep a number of vectors equal to a power of two, that is we fill all indexed slots, the post-selection probability on $\ket{r}$ remains $\frac{1}{2^n}$. On the other hand, when records and centroids are no longer powers of two, the probability further decreases. 
We experimented this aspect by encoding synthetic data in the \textsc{QC3} circuit. In Figure~\ref{fig:postprobR}, we show how the post-selection probability on $\ket{r}$ changes varying the number $M$ of records encoded. We can see that, with $k=2$, it is almost equal to the theoretical one only when $M$ is a power of two. According to the plot, this decrease seems to be proportional to the difference between the number of slots available ($2^m$ where $m=\ceil{\log_2M}$) and the number $M$ of records actually encoded. Instead, when $k=3$, we obtain in general a lower probability but, again, with the peaks when $M$ is a power of two. So, we can claim that the theoretical probability $\frac{1}{2^n}$ is in fact an upper bound of the number of times we measure $\ket{r}$ as 1. Moreover, when the number of records and/or centroids encoded in the quantum circuit is not a power of two, there will be a certain number of slots in the FF-QRAM which remain empty. Thus, when we measure the corresponding qubits $\ket{\kappa}$ and $\ket{m}$, we encounter meaningless configurations (i.e., configurations which do not index vectors loaded in the FF-QRAM). 

As an example, suppose $k=3$. Then, circuit  \textsc{QC3} should have $\hat{k} = \ceil{\log_2k} = \ceil{\log_23} = 2$ qubits for the cluster indexing. This means that one configuration of the basis state of $\ket{\kappa}$ (the last one corresponding to $\ket{11}$) is meaningless, since it does not correspond to any centroid. The same happens when $M$ is not a power of two: in this case there will be a portion of the resulting histogram corresponding to record vectors that actually are not encoded in the circuit. In practice, when  $q_{1:k}$-$k$-Means and $q_{M:k}$-$k$-Means   encounter a meaningless configuration in the measurement results, they do  not consider it for the cluster assignment. Dealing with values for $M$ and $k$ that are not a power of 2, therefore, has a significant impact on the algorithm because all measurements falling into the meaningless configurations will be wasted, leading to poor accuracy.
Flagging only the meaningful index configurations during the data encoding and post-selecting the final measurements based on this flag does not solve this problem, since it is equivalent to classically discard the meaningless configuration on the final histogram. 
 
In light of this, and considering the exponential cost of the post-selection on $\ket{r}$, the number of repetitions one should perform in order to get accurate results becomes huge as the number of features increases and the number of records and/or the number of centroids encoded in the circuit deviates from being a power of two. Since this post-selection probability depends on data distribution, we can consider a preprocessing strategy with the goal of reducing the number of repetitions needed. In~\cite{de2020circuit}, the authors propose a simple preprocessing technique which consists of dividing every entry in the input state (our feature values) by $c = \max_{0 \leq k < N}(\abs{x_k})$. Applying this preprocessing, the post-selection probability becomes $P(\ket{r}=1) = \frac{1}{c^2N}$, and the post-selection success probability is increased by a factor $\frac{1}{c^2}$ where $0<c<1$. However, this strategy can not be used for our algorithm. Indeed, with the FF-QRAM algorithm we can encode any kind of data with feature values in the range $[-1,1]$, but when we deal with non-normalized data, we can not estimate the Euclidean distance between the original data from equation \eqref{pa}. This is because equation \eqref{pa} gives us an estimate of Euclidean distance between quantum states, which are always normalized, and it corresponds to the Euclidean distance between original data only when they are properly normalized. 
For this reason, we can not improve the post-selection probability with the preprocessing technique used in~\cite{de2020circuit}.

The conclusion is that, in general, the number of repetitions each circuit should execute must be proportional to both $k$ and $M$.

\section{Experiments}
\label{sec:experiments}
In this section, we assess the effectiveness of the $q$-$k$-Means algorithms\footnote{Python code at: \url{https://github.com/AlessandroPoggiali/Qkmeans}}. 
We intend to evaluate the capability of the $q$-$k$-Means versions in terms of clustering quality by simulating them on six datasets. 
Since quantum computers currently available are not large enough to test $q$-$k$-Means, we exploit the \textsc{qasm\_simulator} provided by \textsc{qiskit}\footnote{Qiskit library: \url{https://qiskit.org/}}. 

\begin{figure}[t]
\centering
\begin{subfigure}{0.24\textwidth}
    \includegraphics[width=\textwidth]{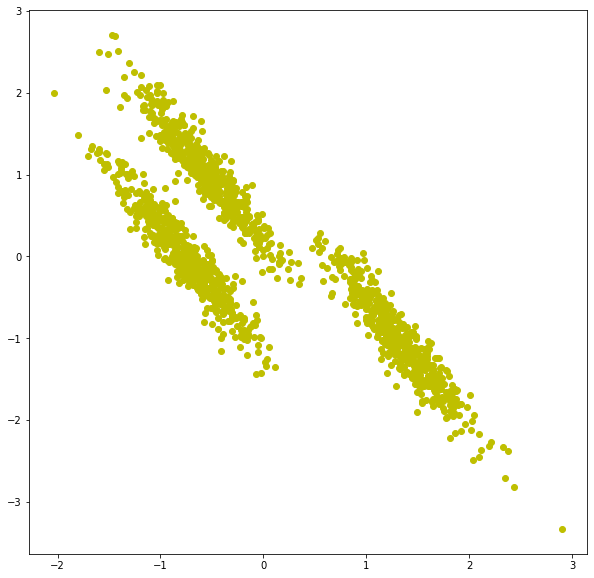}%
    \caption{\textsc{ANISO}}
\end{subfigure}
\begin{subfigure}{0.24\textwidth}
    \includegraphics[width=\textwidth]{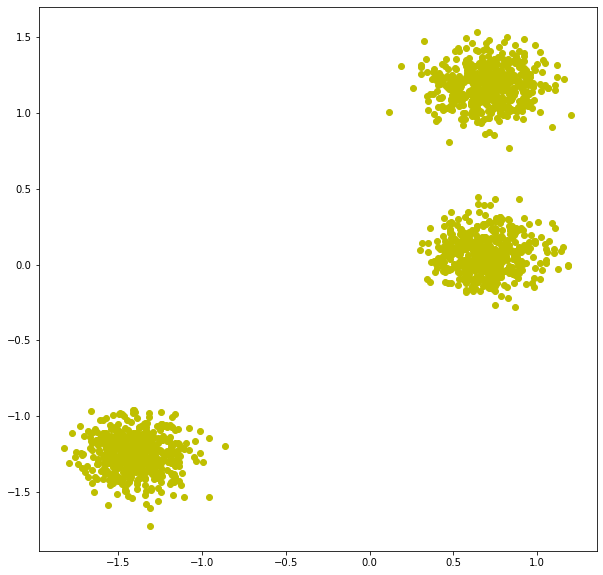}%
    \caption{\textsc{BLOBS}}
\end{subfigure}
\begin{subfigure}{0.24\textwidth}
    \includegraphics[width=\textwidth]{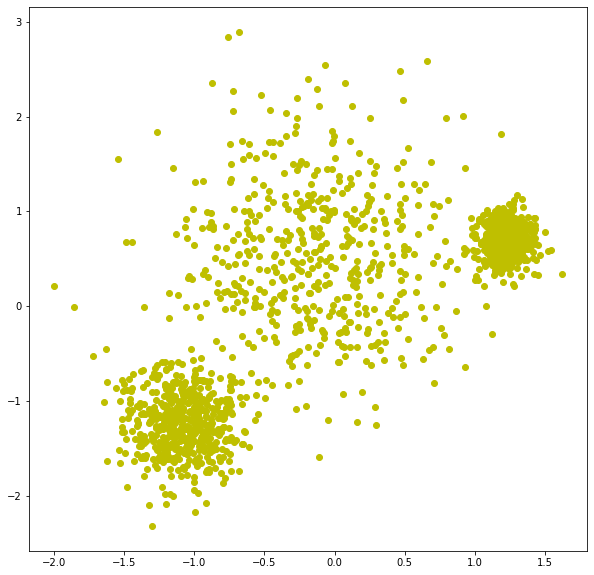}%
    \caption{\textsc{BLOBS2}}
\end{subfigure}
\begin{subfigure}{0.24\textwidth}
    \includegraphics[width=\textwidth]{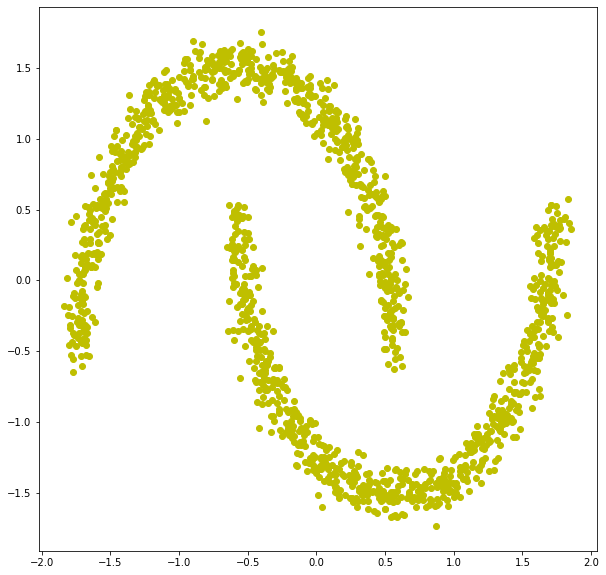}%
    \caption{\textsc{MOON}}
\end{subfigure}

    \caption{Synthetic datasets.}
    \label{fig:synthetic}
\end{figure}

\subsection{Experimental Setting}
\textit{\textbf{Dataset.}}
We consider two groups of datasets: four \textit{synthetic datasets} and two \textit{real datasets}.
The synthetic datasets come from the clustering guide of \texttt{scikit-learn}\footnote{
    \url{https://scikit-learn.org/stable}.
}. These datasets show the characteristics of different clustering algorithms on datasets that are interesting but still in two dimensions. For these datasets, we also have the ground truth (i.e., the actual clustering we want to obtain) so that we can have an objective evaluation of each algorithm. 
They include: \textsc{aniso}, \textsc{blobs}, \textsc{blobs2}, and \textsc{moon} (also known as \textsc{noisymoon}) (Fig. \ref{fig:synthetic}).
The real datasets we take into consideration are \textsc{iris}, and \textsc{wine}, which are also available on \texttt{scikit-learn}. In general, with real datasets, we have no prior knowledge about the clustering result we should achieve. In other words, the number of clusters $k$ has to be selected properly in order to get good clusterization. A very common heuristic used in clustering to determine the number of clusters in a dataset is the \emph{elbow method}~\cite{tan2005introduction}. However, we exploit the target label in \textsc{iris} and \textsc{wine} as ground truth for the ideal clustering, and so we use the actual number of classes (3 in both cases) as value for $k$.

\textit{\textbf{Data Preprocessing.}}
The essential data preprocessing required by $q$-$k$-Means consists of two steps: standardization and normalization with the Inverse Stereographic Projection (\emph{ISP}). 
The standardization of the dataset has the effect of having zero mean and unit variance among all samples. This is common practice in distance-based ML models to prevent features with wider ranges from dominating the distance metric~\cite{mohamad2013standardization}.
The \emph{ISP} normalization applies to each record and allows us to deal with $N$-entries vectors projected on the surface of the unit sphere in the $(N+1)$-dimensional space.
Finally, vector values are converted to suitable angles in order to encode them in the FF-QRAM. 
Moreover, for the real datasets we apply the following features selection: for the \textsc{iris} dataset we select sepal length, petal length and petal width in order to deal with a well-clusterable dataset in three-dimensions; for the \textsc{wine} dataset we select the 7 features with the highest variance.

\medskip
\textit{\textbf{Evaluation Parameters.}}
We experiment with the following parameters to completely understand the behavior of the various algorithms:
\begin{itemize}
\item \textsc{shots} ($t$): number of quantum circuit executions.
\item \textsc{sc\_thresh}: it controls the stopping condition of the algorithm. It is defined as the relative tolerance w.r.t. the Frobenius norm~\cite{golubvanloan} of the difference in the cluster centers of two consecutive iterations. 
\item \textsc{max\_iterations (max\_ite)}: the maximum number of iterations the algorithm can perform.
\item \textsc{M1}: this parameter is meaningful only for $q_{M:k}$-$k$-Means. It denotes how many records we want to load in one single circuit. If $M1$ is smaller than $M$ (the total number of records in a dataset), then several circuits will be created within the same iteration. This allows us to feed the circuit with an arbitrary number of records. Thus, according to Figure~\ref{fig:postprobR}, by setting $M1$ a power of two we achieve the best post-selection probability.
\end{itemize}

\noindent
If not differently specified, we test $q$-$k$-Means with \textsc{max\_ite} $=5$, \textsc{sc\_thresh} $=1.0e-04$, while the number of shots is proportional to the number of records and clusters, and is set as \textsc{shots} $ = 1024$ for $q_{1:1}$-$k$-Means, \textsc{shots} $ = k \times 1024$ for $q_{1:k}$-$k$-Means and \textsc{shots} $ = M1 \times k \times 1024$ for $q_{M:k}$-$k$-Means.

In order to evaluate the algorithm performance and the cluster quality, our analysis considers the following measures. Refer to \texttt{sklearn.metric}\footnote{
    \url{https://scikit-learn.org/stable/modules/classes.html##module-sklearn.metrics}}
    module for the definition of these metrics.
\begin{itemize}
\item \textsc{n\_ite (ite)}: the actual number of iterations $q$-$k$-Means performs. 
\item \textsc{avg\_similarity (sim)}: the concept of similarity is defined in terms of how accurately $q$-$k$-Means assigns the right centroids to records with respect to the classical assignments. 
The \textsc{avg\_similarity} measure is basically the average similarity among all iterations of the algorithm. 
\item \textsc{silhouette (sil)}~\cite{rousseeuw1987silhouettes}: the Silhouette Coefficient measures how an element is similar to its cluster 
with respect to the other clusters. 

\item \textsc{SSE}: is defined as follows
$$
SSE=\sum_{\vec{r}\in M} \|\vec{r}-\vec{c}_{L(\vec{r})}\|^2.
$$
\item \textsc{v\_measure (vm)}~\cite{rosenberg2007v}: it measures the correctness of the clustering assignments with respect to a given ground truth clustering assignment. 
\end{itemize}

\subsection{Results on Synthetic Datasets}

\begin{table}[h!]
\footnotesize
\setlength{\tabcolsep}{0.9mm}
\centering
    \begin{tabular}{c|ccccc|cccccc|cccc}
    \toprule
    & \multicolumn{5}{|c|}{\textsc{$q_{1:1}$-$k$-Means}}&\multicolumn{6}{c|}{\textsc{$\delta$-$k$-Means}}& \multicolumn{4}{|c}{\textsc{$k$-Means}} \\
    \cmidrule{2-16}
    & \textsc{ite} & \textsc{sim} & \textsc{sse} & \textsc{sil} & \textsc{vm}
    & \textsc{$\delta$} & \textsc{ite} & \textsc{sim} & \textsc{sse} & \textsc{sil} & \textsc{vm}
    & \textsc{ite}  & \textsc{sse} & \textsc{sil} & \textsc{vm}\\ \midrule 
    ANISO & 5 & 95.8 & 552.52 & .48 & .62 
                    & 0.3 & 5 & 95.8 & 725.55 & .47 & .52
                    & 5 & 714.75 & .48 & .54\\
    BLOBS & 2 & 100 & 63.48 & .81 & 1.00
                   & 0 & 2 & 100 & 63.47 & .81 & 1.00 
                   & 2 & 63.47 & .81 & 1.00 \\
    BLOBS2 & 5 & 98.5 & 477.43 & .61 & .70
                    & 0.1 & 5 & 98.1 & 436.32 & .62 & .78 
                    & 5 & 435.50 & .63 & .79 \\
    MOON & 1 & 99.6 & 1251.40 & .50 & .38  
                       & 0.1 & 5 & 99.4 & 1250.98 & .50 & .39
                       & 3 & 1250.69 & .50 & .38 \\
    \bottomrule
  \end{tabular}
\caption{$q_{1:1}$-$k$-Means, $\delta$-$k$-Means, and $k$-Means  results on synthetic datasets.
\label{tab:q1kmeans-syn}}
\end{table}

\begin{table}[h!]
\footnotesize
\setlength{\tabcolsep}{0.9mm}
\centering
    \begin{tabular}{c|ccccc|cccccc|cccc}
    \toprule
    & \multicolumn{5}{|c|}{\textsc{$q_{1:k}$-$k$-Means}}&\multicolumn{6}{c|}{\textsc{$\delta$-$k$-Means}}& \multicolumn{4}{|c}{\textsc{$k$-Means}} \\
    \cmidrule{2-16}
    & \textsc{ite} & \textsc{sim} & \textsc{sse} & \textsc{sil} & \textsc{vm}
    & \textsc{$\delta$} & \textsc{ite} & \textsc{sim} & \textsc{sse} & \textsc{sil} & \textsc{vm}
    & \textsc{ite}  & \textsc{sse} & \textsc{sil} & \textsc{vm}\\ \midrule 
    ANISO & 5 & 94.3 & 558.00 & .48 & .61 
                    & 0.5 & 5 & 94.36 & 744.63 & .45 & .51
                    & 5 & 714.75 & .48 & .54\\
    BLOBS & 5 & 99.3 & 70.19 & .80 & .97
                   & 0.4 & 5 & 99.37 & 63.87 & .81 & .99 
                   & 2 & 63.47 & .81 & 1.00 \\
    BLOBS2 & 5 & 97.2 & 494.43 & .59 & .70
                    & 0.3 & 5 & 97.3 & 438.08 & .62 & .79 
                    & 5 & 435.50 & .63 & .79 \\
    MOON & 1 & 99.4 & 1252.09 & .50 & .37 
                       & 0.1 & 5 & 99.4 & 1250.98 & .50 & .39
                       & 3 & 1250.69 & .50 & .38 \\
    \bottomrule
  \end{tabular}
\caption{$q_{1:k}$-$k$-Means, $\delta$-$k$-Means, and $k$-Means  results on synthetic datasets.
\label{tab:q2kmeans-syn}}
\end{table}

Tables~\ref{tab:q1kmeans-syn} and~\ref{tab:q2kmeans-syn} report the evaluation measures observed for $q_{1:1}$-$k$-Means and $q_{1:k}$-$k$-Means respectively, on synthetic datasets, along with the results obtained with $\delta$-$k$-Means and classical $k$-Means on the same configurations.
The values show good performance for both quantum algorithms on \textsc{aniso}, \textsc{blobs}, and \textsc{blobs2}.
In fact, the measures \textsc{sil} and \textsc{vm}  highlight a very good clustering output. 
Meanwhile, in the \textsc{moon} dataset, in spite of having a high similarity values, the \textsc{vm} returned is low. 
This happens because this dataset is not a well-clusterable dataset with algorithms such as $k$-Means that are not able to recognize non spherical clusters.

From Tables~\ref{tab:q1kmeans-syn} and~\ref{tab:q2kmeans-syn} we can also compare the results of the quantum algorithms with the ones obtained with $\delta$-$k$-Means and $k$-Means. The values for $\delta$ are chosen in such a way that $\delta$-$k$-Means similarity is comparable to the $q$-$k$-Means similarity. In this way, we can state that $q$-$k$-Means produces a clustering with error $\delta$. 

\begin{table}[h!]
\scriptsize
\centering
    \begin{tabular}{llrcccc}
    \toprule
    & \textsc{M1} & \textsc{ite} & \textsc{sim} & \textsc{sse} & \textsc{sil} & \textsc{vm} \\ \midrule 
    \multirow{8}{*}{\textsc{ANISO}} & 2 & 5 & 98.40 & 54.39 & .49 & .64 \\
    & 4 & 5 & 97.47 & 54.75 & .49 & .64 \\
    & 8 & 5 & 98.40 & 55.19 & .49 & .64 \\
    & 16 & 5 & 98.00 & 54.31 & .50 & .64 \\
    & 32 & 5 & 98.40 & 54.37 & .50 & .63 \\
    & 64 & 5 & 98.27 & 54.97 & .49 & .62 \\
    & 128 & 5 & 98.53 & 53.93 & .50 & .63 \\
    & 150 & 5 & 97.47 & 55.05 & .50 & .64 \\ \midrule
    \multirow{8}{*}{\textsc{BLOBS}} & 2 & 5 & 98.93 & 8.12 & .77 & .86 \\
    & 4 & 5 & 99.20 & 7.96 & .78 & .95 \\
    & 8 & 5 & 98.67 & 8.82 & .79 & .97 \\
    & 16 & 5 & 99.20 & 7.03 & .77 & .94 \\
    & 32 & 5 & 99.33 & 7.52 & .80 & 1.00 \\
    & 64 & 5 & 98.53 & 7.03 & .79 & .97 \\
    & 128 & 5 & 99.07 & 7.50 & .79 & 1.00 \\
    & 150 & 5 & 97.73 & 8.98 & .77 & .97 \\\midrule
    \multirow{8}{*}{\textsc{BLOBS2}} & 2 & 5 & 94.53 & 47.29 & .62 & .67 \\
    & 4 & 5 & 94.13 & 47.78 & .61 & .69 \\
    & 8 & 5 & 93.87 & 46.66 & .61 & .70 \\
    & 16 & 5 & 92.53 & 54.41 & .60 & .65 \\
    & 32 & 5 & 94.93 & 46.75 & .62 & .69 \\
    & 64 & 5 & 94.00 & 46.42 & .61 & .69 \\
    & 128 & 5 & 93.87 & 47.17 & .62 & .67 \\
    & 150 & 5 & 94.00 & 47.89 & .61 & .68 \\\midrule
    \multirow{8}{*}{\textsc{MOON}} & 2 & 1 & 99.33 & 126.29 & .49 & .38 \\
    & 4 & 1 & 100.00 & 126.38 & .49 & .37 \\
    & 8 & 1 & 98.67 & 126.60 & .49 & .37 \\
    & 16 & 1 & 99.33 & 126.54 & .49 & .35 \\
    & 32 & 1 & 100.00 & 126.39 & .49 & .37 \\
    & 64 & 1 & 99.33 & 126.55 & .49 & .35 \\
    & 128 & 1 & 98.67 & 126.60 & .49 & .36 \\
    & 150 & 1 & 99.33 & 126.54 & .49 & .35 \\\bottomrule
  \end{tabular}
\caption{$q_{M:k}$-$k$-Means results with different $M1$.}
\label{tab:resulttest2}
\end{table}

Concerning the $q_{M:k}$-$k$-Means algorithm, we assess its behavior varying the parameter $M1$, since it can highly determine the accuracy of the final clustering. 
For the sake of the fairness of the experiments, we executed the algorithm on a simplified version of the datasets, sampling at random only 150 points for each of them.  We executed the algorithm varying $M1$ by powers of two up to $M$ ($M1 = M$ included). The results obtained are reported in Table \ref{tab:resulttest2}. As expected, since we chose the rule
for the number of shots ensuring that they were enough for each parameter
configuration, the clustering results obtained are similar regardless of the $M1$ values. This means that $M1$ introduces a trade-off: either we can
have a single quantum circuit ($M1 = M$) that should be executed a significant number of times or we can have many circuits to be executed a lower number of times each. We decided not to report the execution times of the algorithms, since they correspond to simulation times on classical hardware and hence they do not provide a trustful measure of the quantum parallelism. In a real quantum computer we should consider that, even with larger $M1$ values, the total execution time of the circuit remains constant (if we do not consider the cost for data preparation), due to the superposition of the computations. Moreover, increasing $M1$, and hence decreasing the number of different circuits to be passed to the quantum computer, the overhead needed for interacting with the quantum hardware will decrease. 

\begin{table}[t]
\caption{Confusion matrix $k$-Means vs $q$-$k$-Means on \textsc{blobs} dataset.
\label{tab:confusion_matrix_resulttest1}}
\setlength{\tabcolsep}{1.2mm}
\centering
    \begin{tabular}{c|cccc}
    \toprule
    & \textsc{tp} & \textsc{fp} & \textsc{fn} & \textsc{tn}\\ \midrule 
    $q_{1:1}$-$k$-Means & $66.71\%$ & $0\%$ & $0\%$ &  $33.29\%$  \\
    $q_{1:k}$-$k$-Means & $66.36\%$ & $0.35\%$ & $0.35\%$ & $32.94\%$ \\
    $q_{M:k}$-$k$-Means & $65.77\%$ & $1.26\%$ & $1.34\%$ & $31.62\%$ \\
    \bottomrule
  \end{tabular}
\end{table}

A final consideration concerns the confusion matrices in Table~\ref{tab:confusion_matrix_resulttest1}, where we report the percentages of \textit{True Positive (TP)}, \textit{False Positive (FP)}, \textit{False Negative (FN)}, and \textit{True Negative (TN)} for the \textsc{blobs} dataset. 
In this setting, a TP consists in a couple of points which are clustered together by two clustering algorithms under analysis, i.e., in our case $q$-$k$-Means against $k$-Means.
Similarly, a FP consists in a couple of points which are clustered together by $k$-Means but separately by $q$-$k$-Means, assuming $k$-Means is the ground truth. 
A similar reasoning can be done for TN and FN.
In Table~\ref{tab:confusion_matrix_resulttest1} we observe that $q_{1:1}$-$k$-Means outputs the same results as $k$-Means since the value $FP+FN$ for this version is 0. This does not hold for $q_{1:k}$-$k$-Means and $q_{M:k}$-$k$-Means where, instead, the values $FP+FN=0.70\%$ and $FP+FN=2.60\%$ are, however, very small.

\subsection{Results on Real Datasets}

\begin{figure}[t]
\centering
    \includegraphics[width=0.24\textwidth]{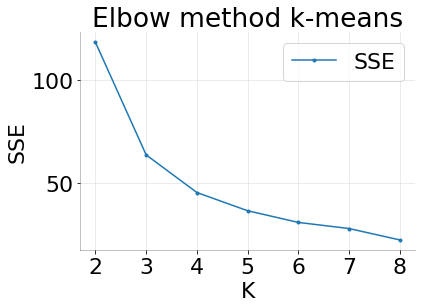}%
    \includegraphics[width=0.24\textwidth]{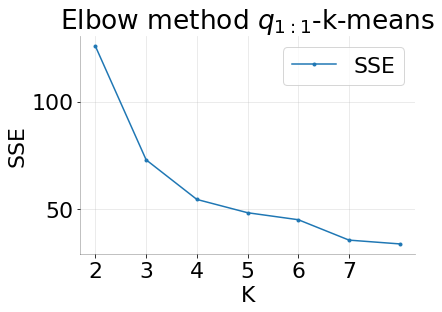}
     \includegraphics[width=0.24\textwidth]{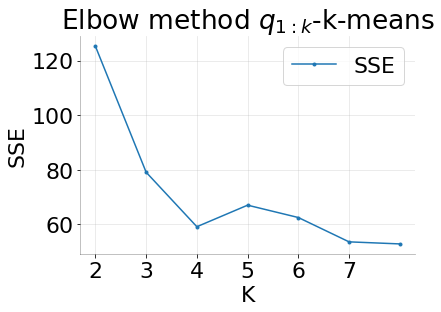}
     \includegraphics[width=0.24\textwidth]{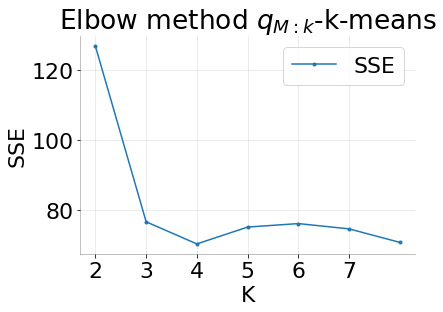}
    \caption{Elbow method on \textsc{iris} dataset with $k$-Means and with the $q$-$k$-Means versions.}
    \label{fig:elbow1}
\end{figure}

\begin{table}[t]
\footnotesize
\setlength{\tabcolsep}{1.2mm}
\centering
    \begin{tabular}{l|rrcrcc|rrcrcc}
    \toprule
    & \multicolumn{6}{|c|}{\textsc{iris}}&\multicolumn{6}{c}{\textsc{wine}} \\
    \cmidrule{2-13}
    & \textsc{$\delta$} & \textsc{ite} & \textsc{sim} & \textsc{sse} & \textsc{sil} & \textsc{vm} & \textsc{$\delta$} & \textsc{ite} & \textsc{sim} & \textsc{sse} & \textsc{sil} & \textsc{vm}\\ \midrule
    $k$-Means & - & 5 & - & 63.56 & .54 & .64 & - & 5 & - & 1085.17 & .25 & .79\\
    \hline
    $q_{1:1}$-$k$-Means & - & 5 & 98.93 & 75.84 & .46 & .67 & - & 5 & 86.74 & 707.31 & .29 & .59\\
    $\delta$-$k$-Means & 0.1 & 5 & 98.73 & 63.66 & .54 & .65 & 5.1 & 5 & 86.07 & 707.53 & .27 & .51\\
    \hline
    $q_{1:k}$-$k$-Means & - & 5 & 97.07 & 77.09 & .46 & .63 & - & 5 & 73.03 & 886.10 & .15 & .32\\
    $\delta$-$k$-Means & 0.2 & 5 & 97.33 & 63.53 & .54 & .65 & 6.3 & 5 & 73.37 & 928.11 & .08 & .28\\
    \hline
    $q_{M:k}$-$k$-Means & - & 5 & 96.04 & 79.73 & .44 & .68 & - & 5 & 78.96 & 859.76 & .18 & .38\\
    $\delta$-$k$-Means & 0.4 & 5 & 96.00 & 64.86 & .52 & .65 & 6.0 & 5 & 78.43 & 752.79 & .24 & .44\\
    \bottomrule
  \end{tabular}
\caption{$q$-$k$-Means vs $k$-Means on \textsc{iris} ($M1 = M = 150$) and \textsc{wine} ($M1 = M = 178$).}
\label{tab:iriswine}
\end{table}

\begin{table}[t!]
\footnotesize
\setlength{\tabcolsep}{0.9mm}
\centering
    \begin{tabular}{r|r|rr|rr|r}
    \toprule
    & \multicolumn{1}{c|}{\textsc{qubits}}&\multicolumn{2}{c|}{\textsc{gates}}& \multicolumn{2}{c|}{\textsc{depth}}&\multicolumn{1}{|c}{\textsc{shots}}\\
    \cmidrule{2-7}
    & & \textsc{high level} & \textsc{transpiled} & \textsc{high level} & \textsc{transpiled} \\ \midrule 
    $q_{1:1}$-$k$-Means & 5 & 53 & 405 & 41 & 307 & 1024 \\
    $q_{1:k}$-$k$-Means & 9 & 111 & 1249 & 83 & 951 & 3072 \\
    $q_{M:k}$-$k$-Means & 23 & 5065 & 118819 & 3064 & 90949 & 460800 \\
    \bottomrule
  \end{tabular}
\caption{Circuit complexities for the IRIS dataset with $k = 3$. For the number of gates and the depth of the circuits we report the qiskit library version (high level) and the compiled version using only the universal quantum gates $R_x$, $R_y$, $R_z$, $P$ and $CNOT$ (transpiled).
\label{tab:circuit}}
\end{table}

Our first test aims at performing the elbow method with the $k$-Means and all the $q$-$k$-Means algorithm versions, separately on the \textsc{iris} dataset, in order to check whether the number of clusters found with this methods corresponds to the desired one ($k=3$). We repeat the execution of these algorithms by varying $k$ from 2 to 8. We obtain the curves in Figure~\ref{fig:elbow1}. As expected, a suitable value of $k$ for all the algorithms is 3 because this is the point where the SSE stops decreasing sharply. The comparison between the classical and the quantum algorithms with this configuration is reported in Table~\ref{tab:iriswine}, while in Table~\ref{tab:circuit} we show the complexities of the circuits in terms of number of qubits, number of gates (both high level and elementary), and depth of the circuits. From this table we observe that, since  we have to load all the data into the QRAM, the number of gates and the depth of the circuit increase with $M$ and $k$. Our empirical rule to choose the number of shots to perform is also proportional to $k$ and $M$, so apparently we do not have a practical advantage in computing the cluster assignment with $q_{1:k}$ and $ q_{M:k}$-$k$-Means. However, in the case all the data are already loaded into the circuit, the algorithms are equivalent. Hence, a more accurate strategy for determining the number of shots needed to achieve a prescribed accuracy should be better investigated in order to improve the cost of $q_{1:k}$ and $ q_{M:k}$-$k$-Means. 

By executing the $q$-$k$-Means algorithms on the $\textsc{wine}$ dataset with $k=3$, we obtain the results in Table~\ref{tab:iriswine}. The results show that $q_{1:1}$-$k$-Means performs better than $k$-Means according to \textsc{SIL} measure, while this does not hold for the $q_{1:k}$-$k$-Means and $q_{M:k}$-$k$-Means versions. However, in general, the \textsc{vm} is quite lower for all quantum versions. A possible explanation can be related to the features selection applied to reduce the number of features from 13 to 7, which is necessary for executing the algorithm in a reasonable amount of time, due to the technological limits. This makes records belonging to different clusters too similar, and the number of shots becomes insufficient to estimate a, therefore affecting the quality of the clustering. Eventually, we observe that the \textsc{vm} of $k$-Means and $q$-$k$-Means on the \textsc{iris} dataset are similar, while in the \textsc{wine} dataset, $k$-Means performs better than $q$-$k$-Means.

\subsection{$q$-$k$-Means on Real Quantum Hardware}

All tests reported up to now were carried out using the \textsc{qasm simulator}, a simulator provided by \textsc{qiskit} which simulates quantum computation by using classical hardware. Here, we show the best we can perform with currently available quantum hardware. IBM offers cloud access\footnote{\url{https://quantum-computing.ibm.com/}} to some of their quantum computers, so it is possible to delegate the execution of a quantum circuit to a real quantum computer. We had access to quantum computers with no more than five qubits. For this reason, we must consider simple instances of $q$-$k$-Means where no more than five qubits are necessary. We take into account for this experiment a small dataset (\textsc{blobs3}) consisting of $M=16$ two dimensional vectors, which form two well-distinguishable spherical clusters. In order to limit the noise introduced by the quantum hardware, we do not apply the \emph{ISP} preprocessing but, instead, we perform clustering in the original features space.

The first test considers $q_{1:1}$-$k$-Means where we use three qubits: one qubit for the ancilla $\ket{a}$, one for the register $\ket{r}$, and one for addressing $N=2$ features (i.e., $\ket{i}$). Instead of simulating each of the $Mk$ quantum circuits locally, we first check the least busy quantum computer available and send to it every circuit to be executed. This requires several steps, like waiting on the queue of a quantum computer and receiving the result, so it introduces an overhead. Furthermore, this communication overhead is paid for each  pair of vectors, so it could highly affect the overall performance of the algorithm. Notice that the number of clusters is not involved in the quantum circuit preparation, but we choose $k=2$ to simplify the overall execution.

\begin{figure}[t]
\centering
    \begin{subfigure}{0.45\textwidth}
    \centering
    \includegraphics[width=0.7\textwidth]{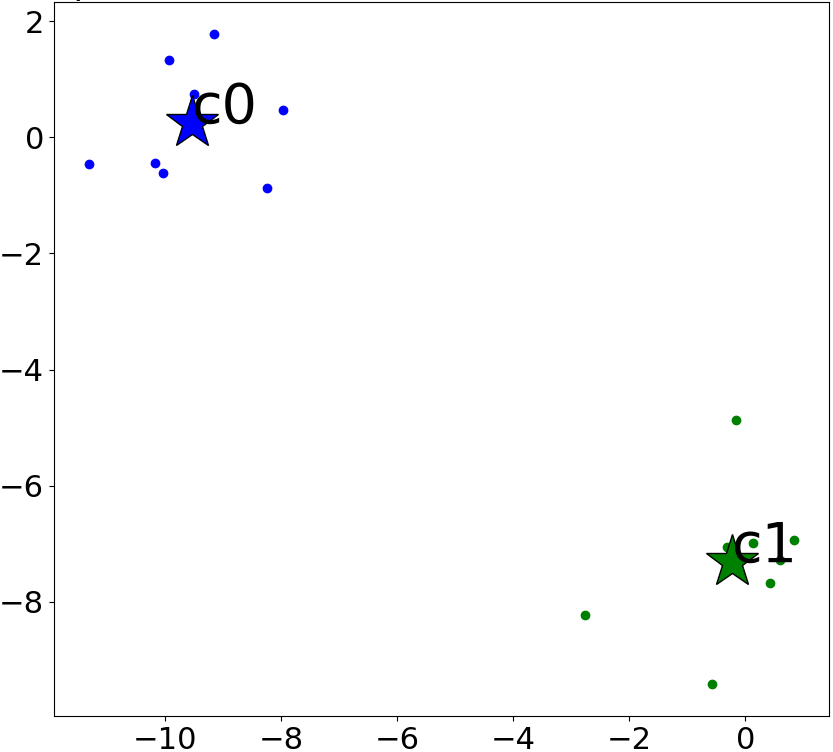}
    \caption{$q_{1:1}$-$k$-Means on quantum hardware.}
    \end{subfigure}
    \begin{subfigure}{0.45\textwidth}
    \centering
    \includegraphics[width=0.7\textwidth]{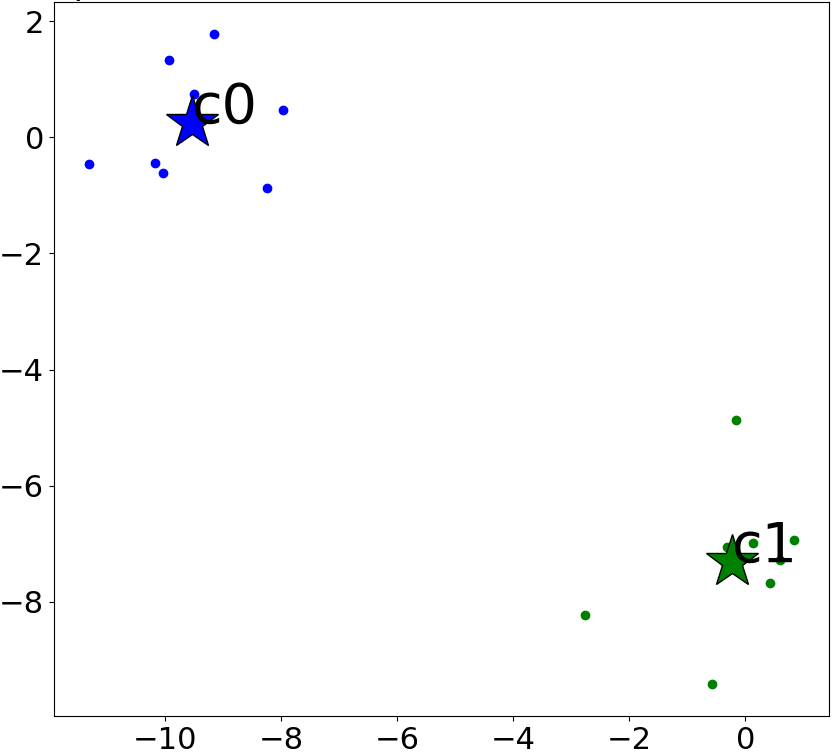}
    \caption{$q_{1:1}$-$k$-Means on simulator.}
    \end{subfigure}
    \begin{subfigure}{0.45\textwidth}
    \centering
    \includegraphics[width=0.7\textwidth]{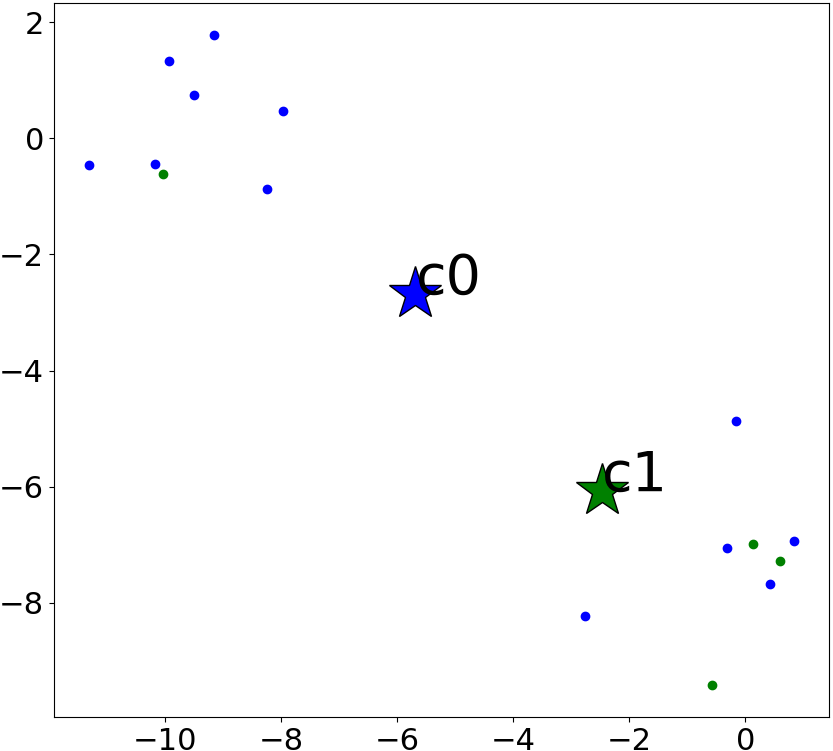}
    \caption{$q_{1:k}$-$k$-Means on quantum hardware.}
    \end{subfigure}
    \begin{subfigure}{0.45\textwidth}
    \centering
    \includegraphics[width=0.7\textwidth]{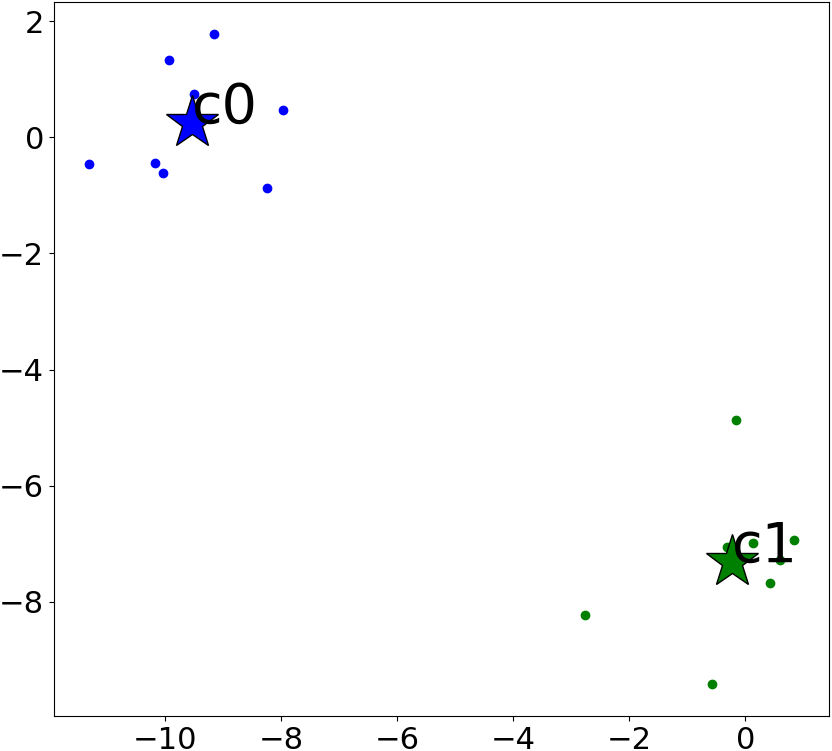}
    \caption{$q_{1:k}$-$k$-Means on simulator.}
    \end{subfigure}
\caption{Clustering result on \textsc{blobs3}.}
\label{fig:blobs3clustering}
\end{figure}

We compare the output of $q$-$k$-Means executed using real quantum hardware with the output of the algorithm executed by the simulator. In Table~\ref{tab:irishw} we report this comparison, while in Figure~\ref{fig:blobs3clustering} we show the clustering obtained.
From Table~\ref{tab:irishw}, we can see that both algorithms succeed in finding the right clusters, with respect to the ground truth, with the same number of iterations. We observe during the test that the time the $q_{1:1}$-$k$-Means algorithm takes in average to execute a single circuit using real hardware is around 6 seconds.

\begin{table}[t]
\footnotesize
\centering
    \begin{tabular}{llccrcc}
    \toprule
    && \textsc{ite} & \textsc{sim} & \textsc{sse} & \textsc{sil} & \textsc{vm} \\ \midrule
    \multirow{2}{*}{$q_{1:1}$-$k$-Means} &
    real hw & 2 & 100 & 89.66 & .79 & 1 \\
    & simulator & 2 & 100 & 89.66 & .79 & 1 \\ \midrule
    \multirow{2}{*}{$q_{1:k}$-$k$-Means} &
    real hw & 5 & 62.5 & 193.72 & .01 & .07 \\
    & simulator & 2 & 100 & 89.66 & .79 & 1 \\ \bottomrule
  \end{tabular}
\caption{$q$-$k$-means: real hardware vs simulator.}
\label{tab:irishw}
\end{table}

The second test takes into account the $q_{1:k}$-$k$-Means where the qubit configuration is arranged in this way: one qubit for the ancilla $\ket{a}$, one for the register $\ket{r}$, one for addressing $N=2$ features ($\ket{i}$), and one for addressing $k=2$ centroids ($\ket{\kappa}$). This is, in fact, the best we can do exploiting the current available quantum hardware. 
We compare the output of $q_{1:k}$-$k$-Means executed using real quantum hardware with the output of the algorithm executed by the simulator. As before, we report in Table~\ref{tab:irishw}  this comparison, and we show  in Figure~\ref{fig:blobs3clustering} the clusters obtained.
The bad results obtained with $q_{1:k}$-$k$-Means are due to the noise present in quantum computers. In this case there is too much uncertainty in determining the right cluster for the records. 
In this case, we observe that the algorithm takes a time similar to the previous case (around 6 seconds) to execute each quantum circuit on real hardware. This proves that, even though the circuit now contains more vectors, the actual computation takes place in superposition and so the generalization introduced does not affect the execution time of the single circuit.
To conclude, until a large-scale noise-free quantum computer is available, testing complex quantum circuits on real quantum hardware will be an unfeasible task.

\section{Conclusion}
\label{sec:conclusion}
In this paper, we have proposed $q$-$k$-Means, a hybrid approach for clustering classical data. 
We provided three versions of the algorithm which implement different quantum subroutines to boost the \textit{cluster assignment step} of the classical $k$-Means. The first version, $q_{1:1}$-$k$-Means computes in a quantum way the Euclidean distance between two $N$-dimensional vectors, i.e., a record and a cluster centroid. The complexity of this step is $O(Mk)$ per circuit execution, where $M$ and $k$ are the number of records and the number of centroids, respectively. The $q_{1:k}$-$k$-Means version executes quantum circuits able to assign the right cluster to a record according to its closeness to all centroids, leading the complexity of the \textit{cluster assignment step} to $O(M)$ per circuit execution. The final version of the algorithm, $q_{M:k}$-$k$-Means, is a further generalization where the entire step of cluster assignment is accomplished by executing only one quantum circuit exploiting quantum parallelism. In this way, the cost of the \textit{cluster assignment step} becomes $O(1)$ per circuit execution.
In this paper, we do not estimate mathematically the number of shots that would guarantee the accuracy of the results, while, instead, it has been determined empirically. As we already observed, with the current choice, where the number of repetitions of the circuit increases linearly with the number of records and clusters, we do not see any advantage in using algorithms $q_{1:k}$-$k$-Means and $q_{M:k}$-$k$-Means over the $q_{1:1}$-$k$-Means. We plan to further investigate this issue to see if we can reduce the number of shots keeping the advantage of a method  maximizing the quantum parallelism.

The experiments have shown that the three versions produce comparable clustering results, as long as the number of shots is properly chosen.
Finally we have found out that the currently quantum hardware is not powerful enough to deal with even simple instances of our algorithm. In fact, only $q_{1:1}$-$k$-Mean was able to produce good results using a real quantum computer but the time it takes is prohibitive for everyday use due to the huge communication overhead.

Future works include designing other techniques which aim to reduce or eliminate the post-selections needed, to improve the quantum algorithm performance.

\section*{Acknowledgments}
This work is supported by Universit\`a di Pisa under the ``PRA – Progetti di Ricerca di Ateneo'' (Institutional Research Grants) - 
Pr. no. PRA\_2020-2021\_92, 
``Quantum Computing, Technologies and Applications'',
 by GNCS-INdAM and  by the European Community Horizon~2020 programme under the funding schemes: H2020-INFRAIA-2019-1: Research Infrastructure G.A. 871042 \textit{SoBigData++}.

\bibliographystyle{elsarticle-num} 
\bibliography{biblio}





\end{document}